\documentclass[twocolumn]{article}

\usepackage{appendix}
\usepackage{color}
\usepackage{subcaption}
\usepackage{multirow}
\usepackage{multicol}

\usepackage{PRIMEarxiv}

\usepackage[utf8]{inputenc} 
\usepackage[T1]{fontenc}    
\usepackage{hyperref}       
\usepackage{url}            
\usepackage{booktabs}       
\usepackage{amsfonts}       
\usepackage{nicefrac}       
\usepackage{microtype}      
\usepackage{lipsum}
\usepackage{fancyhdr}       
\usepackage{graphicx}       
\graphicspath{{media/}}     
\usepackage[table]{xcolor}
\usepackage{abstract} 

\hypersetup{
    colorlinks=true,
    linkcolor=blue,
    filecolor=blue,      
    urlcolor=blue,
    citecolor=cyan,
}

\title{
Self-Reference Deep Adaptive Curve Estimation for Low-Light Image Enhancement
}

\author{
  Jianyu Wen\\
  School of Mechanical Engineering and Automation\\
  Harbin Institute of Technology, Shenzhen\\ China \\
  \texttt{22s153130@stu.hit.edu.cn} \\
  \And
  Chenhao Wu\\Maynooth International Engineering College\\
  National University of Ireland Maynooth\\ Ireland\\
  \texttt{CHENHAO.WU.2020@MUMAIL.IE} \\
  \AND
  Tong Zhang\\
  Department of Computer Science and Engineering\\
  Southern University of Science and Technology\\China \\
  \texttt{11911831@mail.sustech.edu.cn} \\
  \And
  Yixuan Yu\\
  College of Sciences\\
  \qquad \qquad Northeastern University\qquad \qquad \quad\\ China \\
  \texttt{20201910@stu.neu.edu.cn} \\
  \And
  Piotr Świerczyński\\
  NODAR Inc. \\USA\\
  \texttt{pwswierczynski@gmail.com} \\
}

\begin{document}
\twocolumn[
\begin{@twocolumnfalse}
\maketitle
\end{@twocolumnfalse}
]

\begin{abstract}
   
   In this paper, we propose a 2-stage low-light image enhancement method called Self-Reference Deep Adaptive Curve Estimation (Self-DACE). In the first stage, we present an intuitive, lightweight, fast, and unsupervised luminance enhancement algorithm. The algorithm is based on a novel low-light enhancement curve that can be used to locally boost image brightness. We also propose a new loss function with a simplified physical model designed to preserve natural images' color, structure, and fidelity. We use a vanilla CNN to map each pixel through deep Adaptive Adjustment Curves (AAC) while preserving the local image structure. 
   Secondly, we introduce the corresponding denoising scheme to remove the latent noise in the darkness. We approximately model the noise in the dark and deploy a Denoising-Net to estimate and remove the noise after the first stage. 
   Exhaustive qualitative and quantitative analysis shows that our method outperforms existing state-of-the-art algorithms on multiple real-world datasets. Codes can be found here - \href{https://github.com/John-Venti/Self-DACE}{https://github.com/John-Venti/Self-DACE}.
   
\end{abstract}

\begin{figure}[t]
  \centering
    \begin{subfigure}{0.49\linewidth}
    \includegraphics[width=\linewidth]{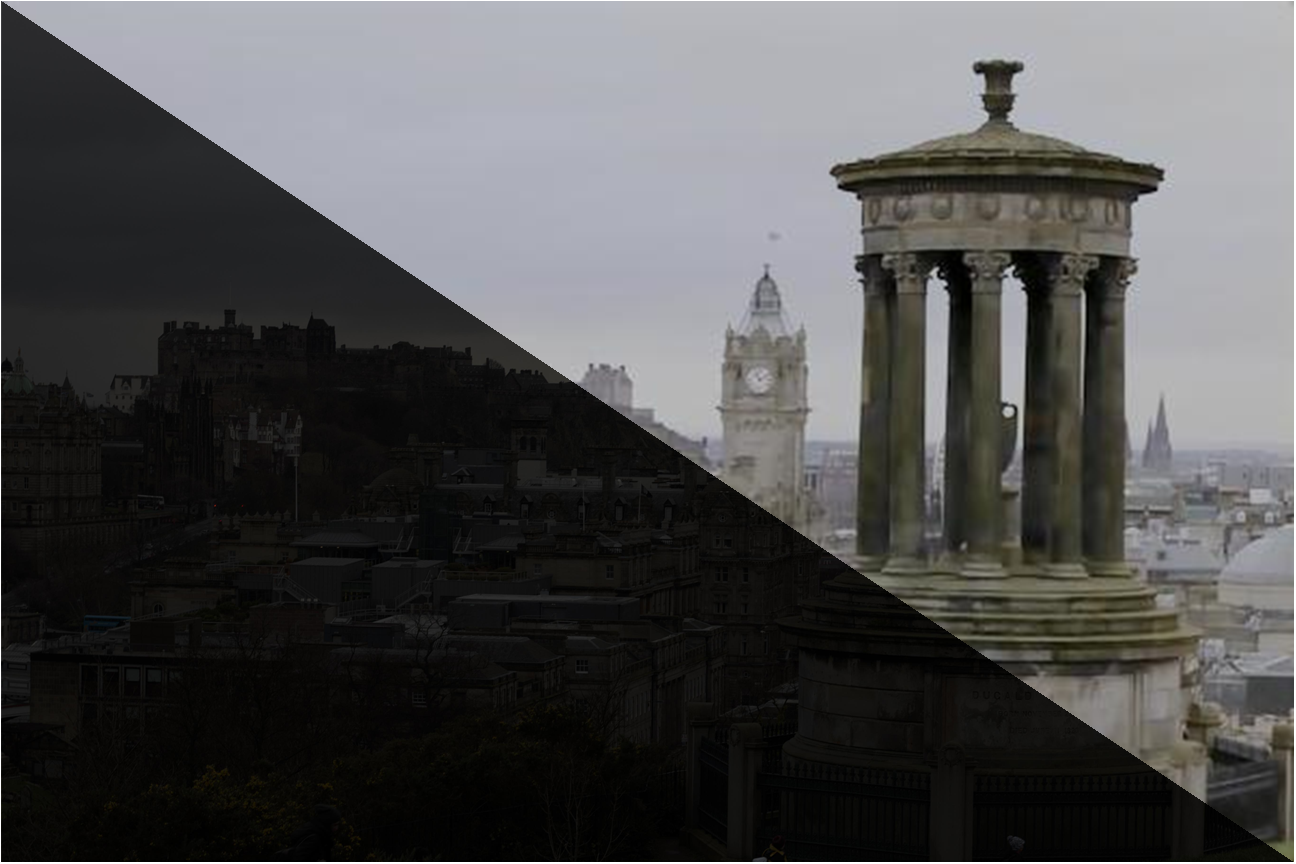} 
  \end{subfigure}
  \hfill
  \begin{subfigure}{0.49\linewidth}
    \includegraphics[width=\linewidth]{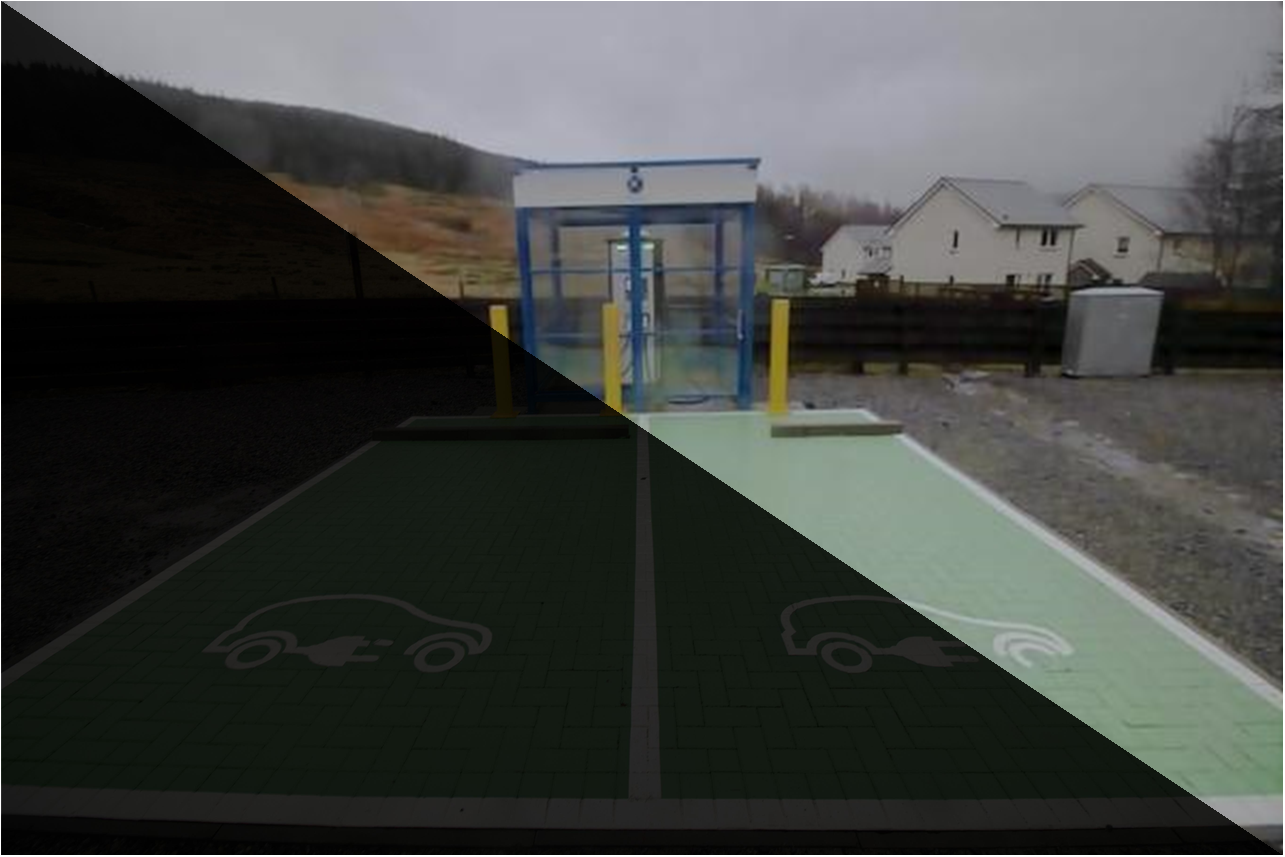}
  \end{subfigure}
  \hfill
  \begin{subfigure}{0.49\linewidth}
    \includegraphics[width=\linewidth]{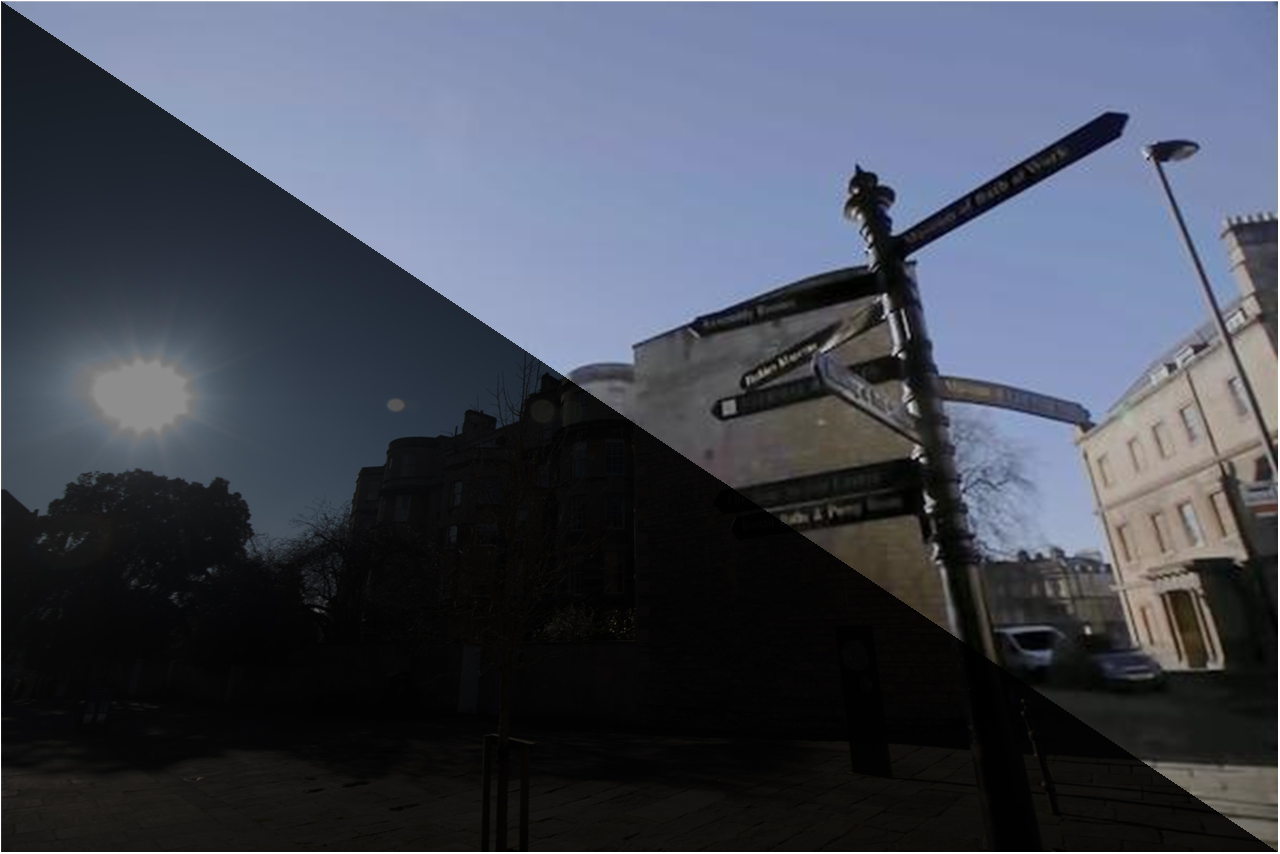}
  \end{subfigure}
  \hfill
  \begin{subfigure}{0.49\linewidth}
    \includegraphics[width=\linewidth]{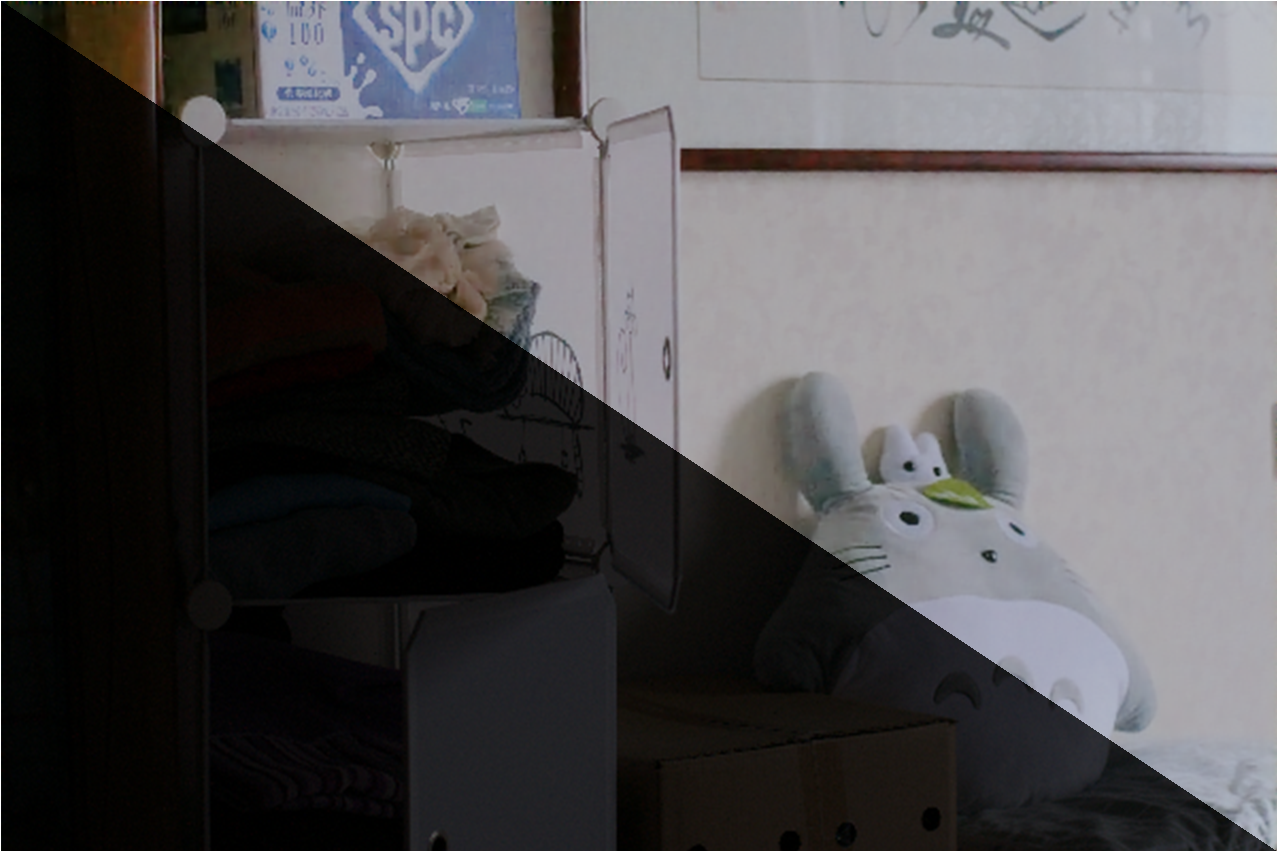} 
  \end{subfigure}
  \caption{Visual comparison with original low-light image on LOL and SCIE dataset. The enhanced images of our method are on the top-right corners. Input low-light images are on the bottom-left corners.}
  \label{fig:example}
\end{figure}

\section{Introduction}

Images taken in low-light conditions are now ubiquitous with the increased popularity of smart devices and the broader adoption of computer vision in everyday applications. Despite the growing interest of the research community in the field of low-light image enhancement, there does not exist a single golden standard technique. Such a technique would have to satisfy several challenging criteria simultaneously: it would have to be fast and not require extensive computational resources to allow for the use by edge devices common in robotics, consumer electronics, and the Internet of Things. For broad applicability across diverse domains and cameras, the solution should be amenable to unsupervised training without reliance on labeled datasets. Acquiring images in sufficiently diverse ground truth lighting is virtually unattainable in real-world scenarios~\cite{2021survey}, especially in complex urban environments.

Mainly inspired by ZeroDCE~\cite{guo2020zero} algorithm, we develop a novel  Self-Reference Deep Adaptive Curve Estimation (Self-DACE) low light image enhancement method. We show that employing a more flexible class of adaptive adjustment curves iteratively leads to an enhancement method applicable to complex illumination scenarios with a wider dynamic range. Our main contributions are as follows:

\begin{itemize}
    \item We propose a novel unsupervised and self-reference loss function rooted in Retinex illumination theory and an estimation factor approximating the physical model of the camera. This loss function is primarily aimed at curve-based image enhancement, but it can also be applied to other neural network-based methods.
    \item  We design an iterative low-light enhancement method using a new adjustment curve. Our flexible and parameterized adjustment curves can enhance images locally without introducing undesirable artifacts.
    \item We propose a lightweight convolutional network suitable for application of luminance enhancement in constrained environments typical for edge devices. The network can be used in real-time image processing scenarios, such as helping mobile robots perceive their surroundings in poorly-lit environments.
    \item We develop an unsupervised denoising method to remove amplified noise hidden in the darkness brought about by luminance enhancement.
    \item We perform extensive qualitative and quantitative analysis of the algorithm's performance to show excellent generalizability among several diverse datasets. We compare it to other state-of-the-art methods using several measures to evaluate the visual performance. 
\end{itemize}

The remainder of this paper is organized as follows: Section 2 provides an overview of related work. Section 3 introduces our algorithm of 2 stages, including a new adjustment curve, the novel loss function, and the details of neural network architecture. We analyze the performance of our method in Section 4 and show an ablation study demonstrating the importance of each component of the loss function and effect of the network parameters. We then compare our algorithm with leading methods across several datasets and metrics.  Finally, we conclude this paper in Section 5.

\label{sec:intro}

\section{Related Work}
Low-light enhancement algorithms can be divided depending on their core methodology. Adjustment methods use local operators to adjust the original low-light images. Reconstruction approaches use convolutional networks to reconstruct entire input images, often incorporating other image enhancement tasks such as denoising.

{\bf Adjustment methods.}
Utilizing  tunable enhancement curves is one of the primary approaches used in low-light image enhancement of the first class. Such methods offer simplicity and preserve image fidelity, minimizing visual artifacts.

In the beginning, using the image's wavelet decomposition is proposed~\cite{loza2013automatic}  to create an enhancement function modeled as bi-variate Cauchy distribution. The algorithm offers contrast improvement in low-light and unevenly illuminated scenes.
ExCNet~\cite{zhang2019zero} is a zero-shot learning method based on S-curve adjustment. It includes spatial consistency in the loss function to improve the convergence of the S-curve parameters.
Then ZeroDCE and ZeroDCE++ algorithms \cite{guo2020zero,li2021learning} are developed, which use iterative approximations of a $\gamma$ function through quadratic polynomials to solve the problem by applying pixel-wise $\gamma$ enhancement function to poorly illuminated images. Though effective in most of time, loss design is so simplistic that it can't handle complex lighting situations.
Later, combining denoising and low-light image enhancement is proposed \cite{wang2022image}, extending the concept of ZeroDCE.
Then a deep reinforcement learning algorithm using Markov processes to incorporate semantic information in low-light enhancement\cite{zhang2021rellie} is developed. 
A semantic-guided zero-shot method\cite{zheng2022semantic}  is also used to low-light enhancement.

Several methods explicitly employ Retinex theory~\cite{land1977retinex} to find an illumination map and enhance low-light images. 
LIME~\cite{Lime2017} combines per-channel illumination initialization with a prior description of  image structure.
RUAS~\cite{liu2021retinex} is a pretty lightweight but effective enhancement network for low-light images using a cooperative reference-free learning strategy to discover an optimal neural network architecture.
SCI learning framework~\cite{ma2022toward} builds on top of RUAS and involves a cascaded illumination learning process with weight sharing to accelerate the algorithm and further improve its results.

{\bf Reconstruction methods.}
Methods based on image reconstruction utilizing a specially designed network have been broadly applied to image enhancement problems, including low-light enhancement.

JED~\cite{ren2018joint} and Retinex-Net~\cite{wei2018deep} use decomposition inspired by the Retinex model for simultaneous low-light enhancement and noise reduction. 
This idea is extended further by RRDNet~\cite{zhu2020zero}, which is a CNN with three separate branches for image noise, illumination, and reflectance.
KinD~\cite{zhang2019kindling} takes a different approach, removing the need for ground-truth labels by designing a neural network trained using pictures of the same scene but with varying illumination.
SGM-Net~\cite{yang2021sparse} proposes a mapping network based on Retinex-Net for correcting underexposed images.
Using bright channel prior (BCP) for LLIE\cite{lee2020unsupervised} suffers from amplified noise in the processed images. 
Later, Uretinex-Net~\cite{wu2022uretinex}  builds denoising and low-light image enhancement as an optimization problem, while LLFlow~\cite{wang2022low} constructs a probabilistic flow for the tasks.
They both struggle with over-fitting on the training data and tend to produce significant artifacts.
Context-Sensitive Decomposition Network (CSDNet)\cite{ma2021learning} is proposed to exploit scene-level contextual dependencies on multiple scales.

Generative adversarial networks (GANs) have shown promising results when applied to image enhancement tasks, including low-light enhancement~\cite{wang2019rdgan,shi2019low}.
EnlightenGAN~\cite{jiang2021enlightengan} realizes image restoration and enhancement, when there is a lack of paired training data.
Moreover, LeGAN~\cite{fu2022gan} addresses the problems of noise and color bias by incorporating an illumination-aware attention module in the network architecture to improve the feature extraction process.

\begin{figure}[t]
  \centering
    \begin{subfigure}{0.49\linewidth}
    \includegraphics[width=\linewidth]{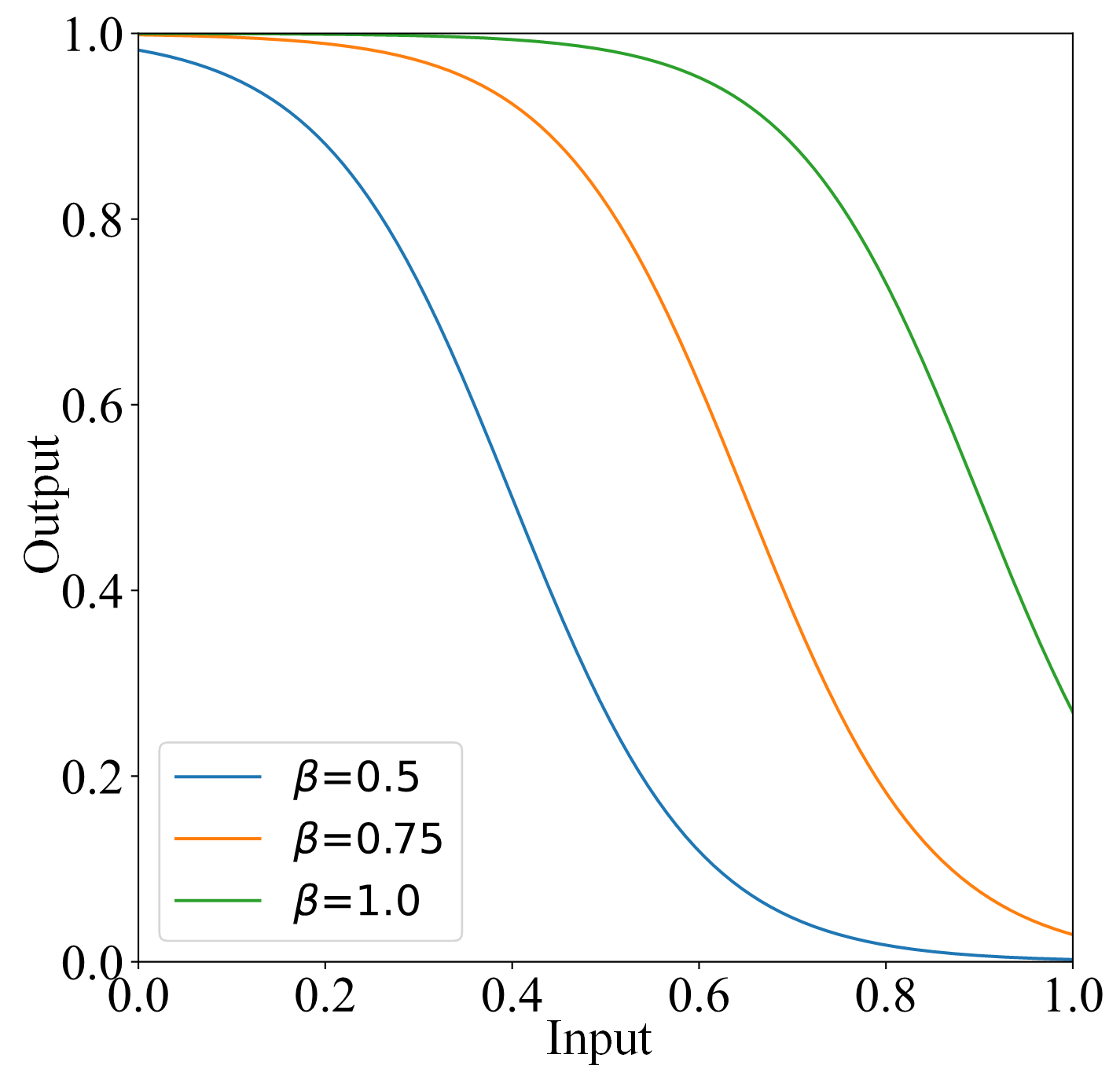} 
    \caption{}
    \label{fig:sigmoid}
  \end{subfigure}
  \hfill
  \begin{subfigure}{0.49\linewidth}
    \includegraphics[width=\linewidth]{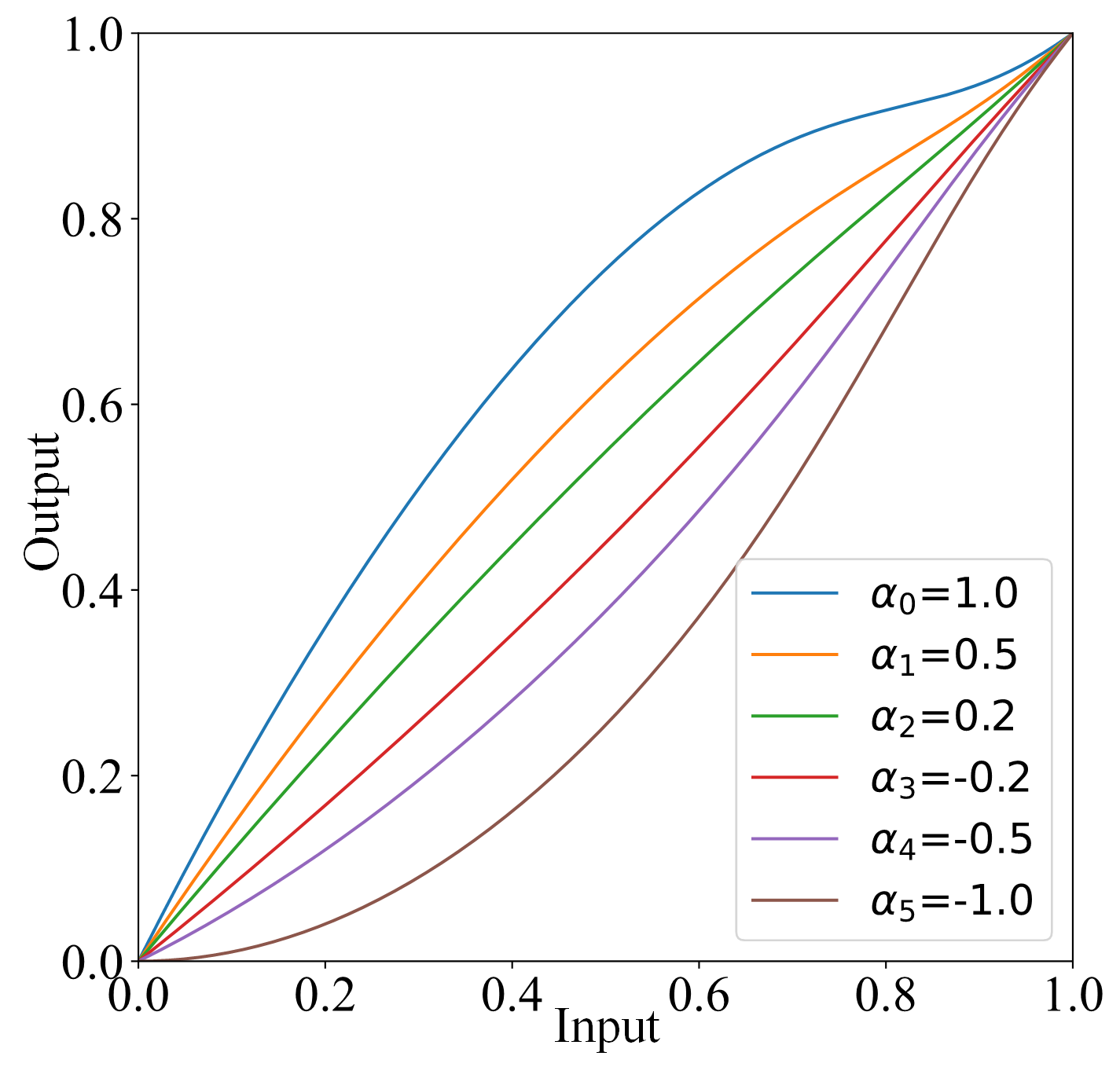} 
    \caption{}
    \label{fig:alpha}
  \end{subfigure}
  \hfill
  \begin{subfigure}{0.49\linewidth}
    \includegraphics[width=\linewidth]{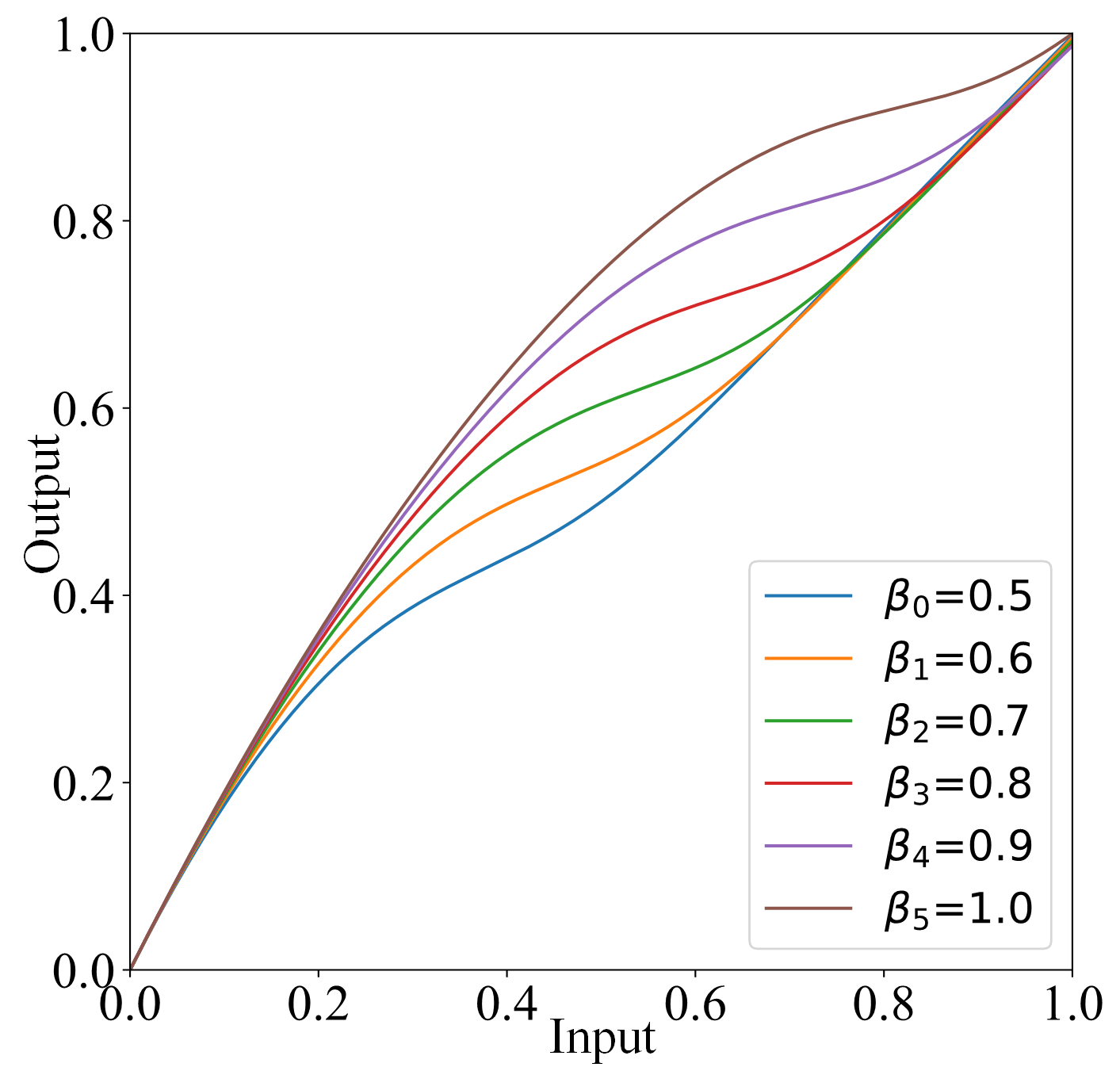}
    \caption{}
    \label{fig:beta}
  \end{subfigure}
  \hfill
  \begin{subfigure}{0.49\linewidth}
    \includegraphics[width=\linewidth]{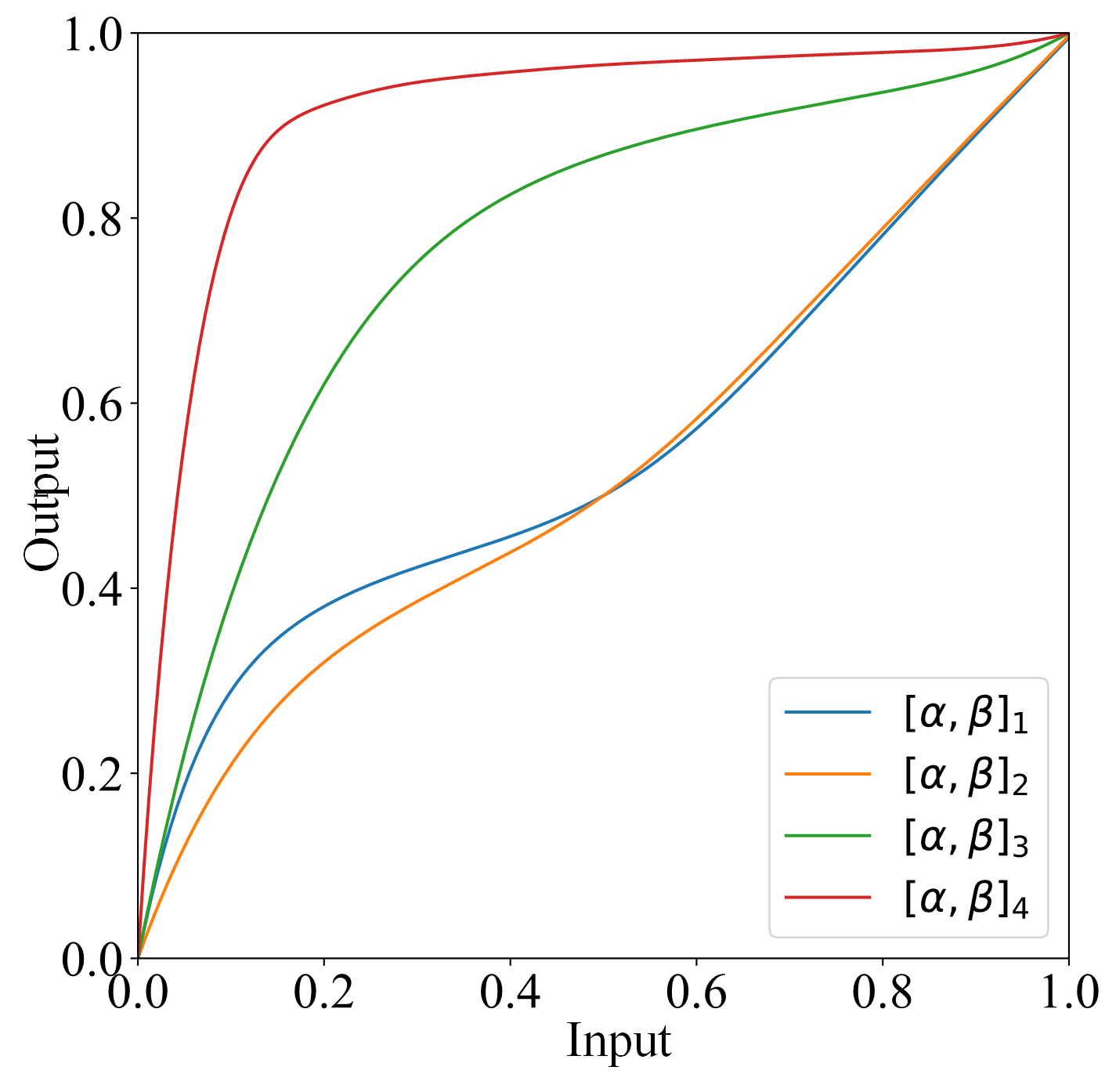} 
    \caption{}
    \label{fig:fusion}
  \end{subfigure}
  \caption{(a) shows the shape of $S(\beta;I^c)$ depending on the choice of parameter $\beta$. (b) and (c) show the influence of parameters on curves when $\beta=1$ and $\alpha=1$ respectively. In (d), $[\alpha,\beta]_1=[0.5,0.5]$, $[\alpha,\beta]_2=[0.3,0.5]$, $[\alpha,\beta]_3=[0.5,1]$ and $[\alpha,\beta]_4=[1,1]$,  are respectively used to adjust the low-light image 4 times iteratively.}
  \label{fig:curve}
\end{figure}

\begin{figure}[t]
  \centering  
  \includegraphics[width=\linewidth]{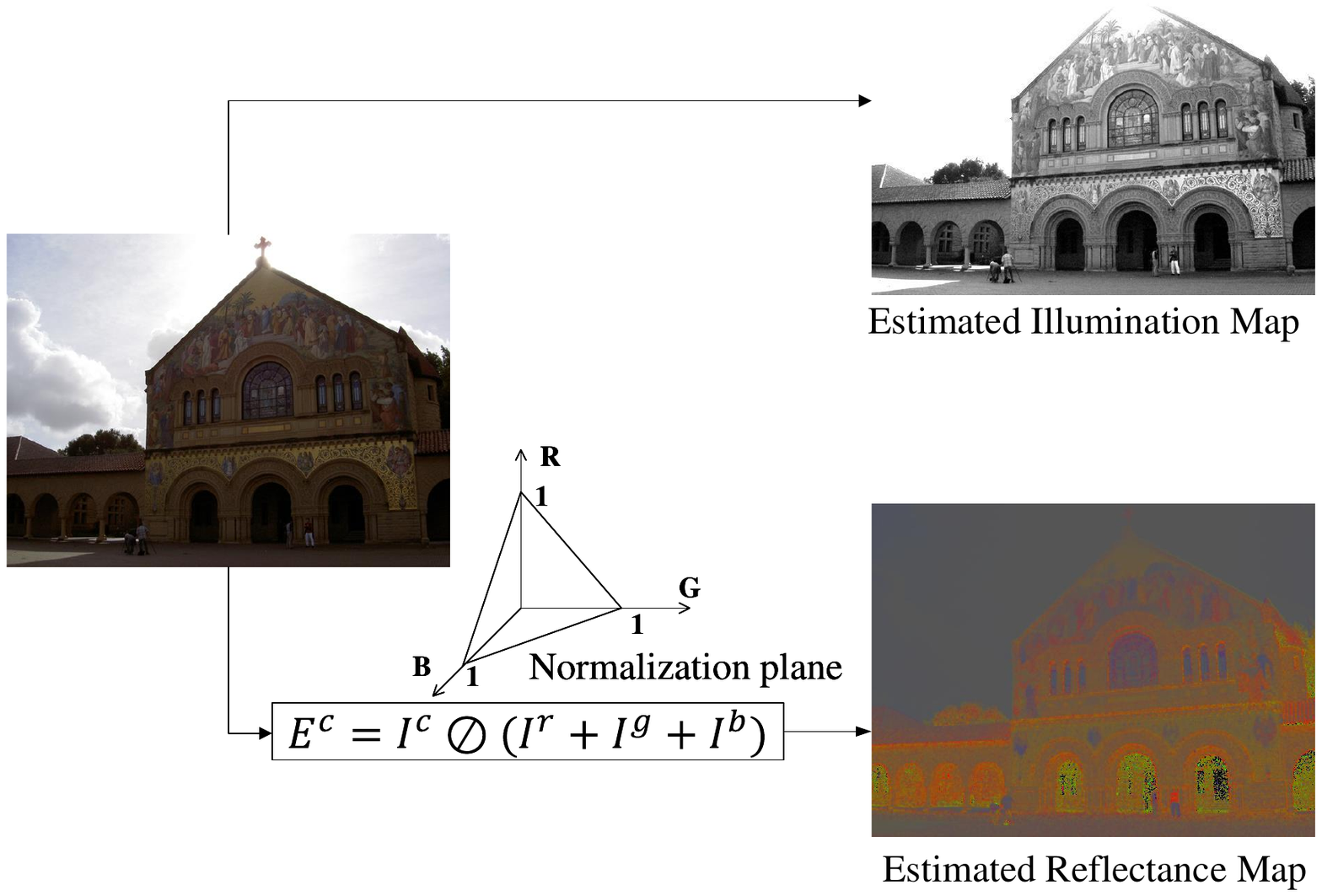} 
  \caption{The Estimated Retinex decomposition. The estimated reflectance map can be represented by an image consisting of estimators on 3 channels. The estimated illumination map consequently can be illustrated by a gray-scale image, where the intensity value of each pixel represents the light intensity.}  
  \label{fig:est}  
\end{figure}

\begin{figure*}[t]
  \centering
  \includegraphics[width=\linewidth]{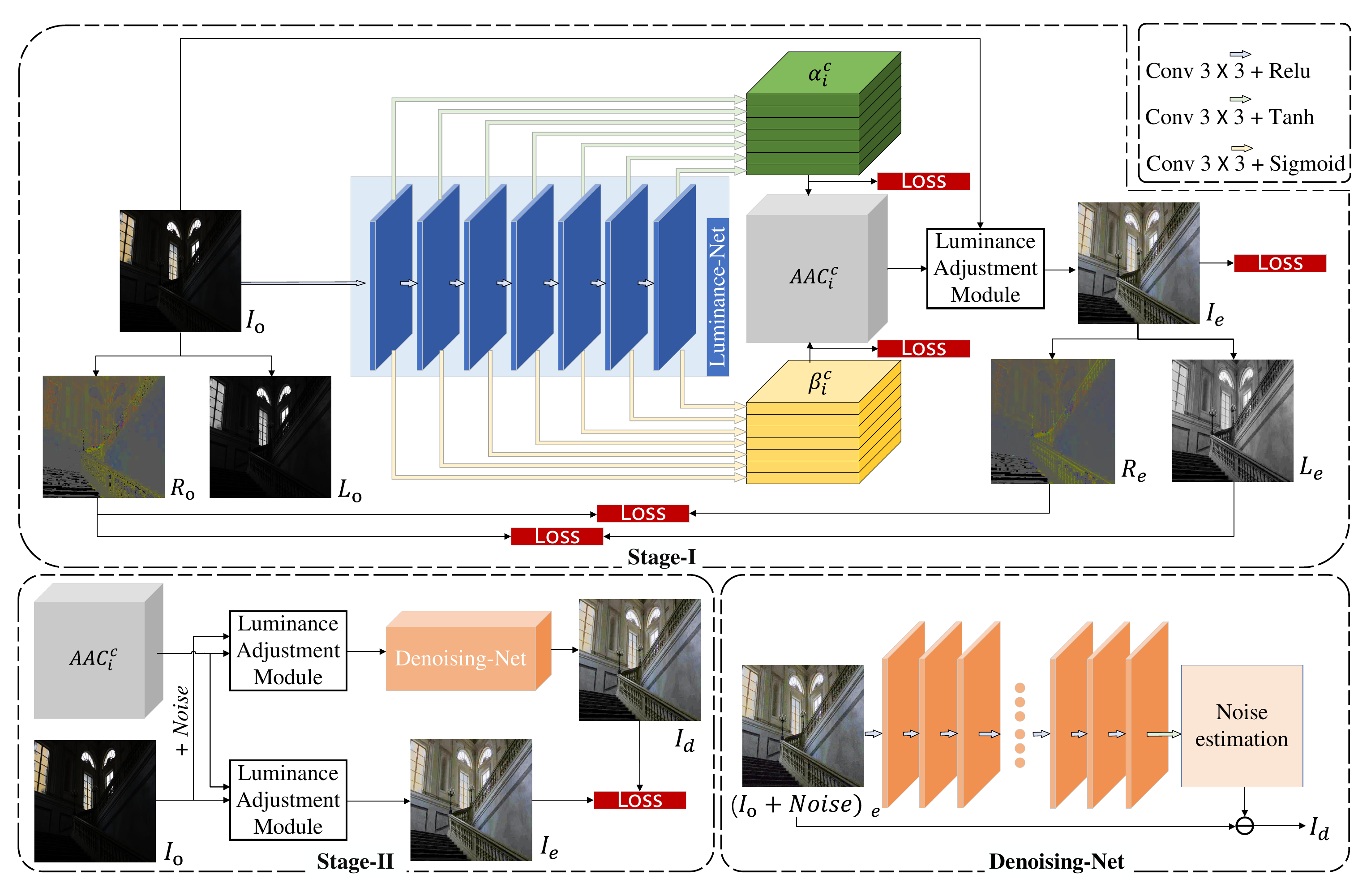} 
  \caption{Stage-I obtains the estimated optimal parameters of AAC from each layer of Luminance-Net, a vanilla CNN. AAC is then used to enhance the input image. Stag-II uses Denoising-Net to estimate the noise after luminance adjustment to get the final result.} 
  \label{fig:frame}  
\end{figure*}

\section{Methods}
In this section, we introduce Adaptive Adjustment Curves (AAC) applied iteratively to each image pixel. We detail the estimation factor from Retinex theory, central to our novel self-reference and unsupervised loss functions. We propose a two-stage framework for dark luminance adjustment and latent noise removal.

\begin{figure*}[t]
  \centering
   \includegraphics[width=1\linewidth]{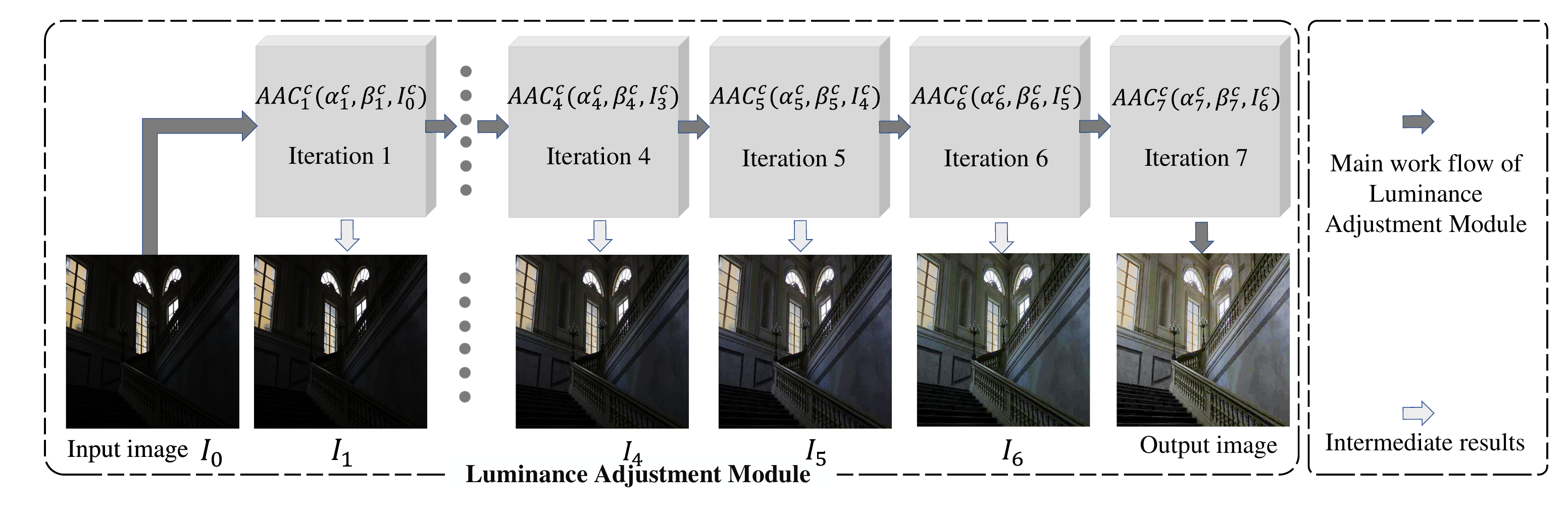}
   \caption{The main process of Luminance Adjustment Module. The low-light image input is enhanced by AAC iteratively. The parameters of $AAC^c_i$ are the outputs of Luminance-Net.}
   \label{fig:iter}
\end{figure*}

\subsection{Adaptive Adjustment Curves}
Following the approach initially proposed in~\cite{guo2020zero}, we aim to design a flexible adjustment curve, which can handle images in complex lighting without artifacts and overexposure. We propose a monotonous, differentiable function mapping a $[0, 1]$ interval onto itself. The adjustment curve can be expressed as
\begin{equation}
  AAC{^c_i}(\alpha{^c_i},\beta^c_i; I^c) =  I^c + \alpha^c_i \otimes \frac{1}{\beta^c_i} \otimes S(\beta^c_i;I^c) \otimes I^c\otimes(\beta^c_i - I^c),
  \label{eq:AAC}
\end{equation}
where
\begin{equation}
  S(\beta{^c_i};I^c)=Sigmoid(-I^c+\beta{^c_i}-0.1), c \in \lbrace r, g, b \rbrace.
  \label{eq:Sig}
\end{equation}

$I^c$ denotes the per-pixel intensity value of $c$-channel normalized to $[0, 1]$, $c$ denotes color channels in the RGB color space, and $\otimes$ represents element-wise multiplication. $\alpha(\mathbf{x}) \in [-1, 1]$ and $\beta(\mathbf{x})\in[0.5, 1]$ are trainable pixel-wise scalar maps. Parameter $\alpha$ adjusts the magnitude of AAC, whereas $\beta$ controls the range of enhancement. We constrain $\beta$ to values inside of the interval $[0.5, 1]$, which improves the training convergence.
We summarize the effect the choice of parameters has on the form of the enhancement curve in Figures~\ref{fig:alpha} and~\ref{fig:beta}.

We introduce $S(\beta;I_c)$ in Figure~\ref{fig:sigmoid} to suppress parts of $I^c$ with intensity exceeding $\beta$, at the same time enhancing the low-intensity levels lower than $\beta$. Subtracting 0.1 can help $S(\beta;I_c)$ suppress AAC in advance to avoid curve oscillation. 

By iteratively applying AAC, the final curve can accurately approximate the behavior of known adjustment curves, for example, S-curve and $\gamma$ curve, see Figure~\ref{fig:fusion}. The flexibility of AAC leads to excellent performance in low-light image enhancement, resulting in very few artifacts. 

\subsection{Estimation Factor}
According to Retinex theory, every image can be decomposed into a product of an illumination map and a reflectance map. The reflectance map represents the image's inherent color distribution, independent of illumination, and the reflectance map is constant under illumination changes.

Existing methods often use the maximum value~\cite{Lime2017,wu2022uretinex} or the original image~\cite{liu2021retinex,ma2022toward} to estimate the reflectance map. We take a normalization approach accounting for the intensity of all three color channels, and define the reflectance decomposition as
\begin{equation}
  E^c = I^c\oslash (I^r + I^g + I^b + \varepsilon), \quad \varepsilon > 0,
  \label{eq:factor}
\end{equation}
where $\oslash$ stands for pixel-wise division. We add a small offset $\varepsilon = 10^{-4}$ to avoid division by $0$.

Assuming white incident light captured as 3 color channels, the estimation closely represents the camera-captured image's reflectance under simplified conditions. This factor normalizes input by projecting all pixels onto a normalization plane, and enabling the estimation of the illumination and reflectance maps, see Figure~\ref{fig:est}.

\subsection{Framework}
We perform luminance enhancement and denoising in two stages to address challenging noise in dark images, outlined in Figure~\ref{fig:frame}.

In Stage-I, we deploy a 7-layer vanilla CNN with 32 convolutional $3\times3$ kernels, as Luminance-Net to handle the low-light image input. Since CNNs capture rich hierarchical features from shallow to deep layers~\cite{liu2021densernet}, the $i$th iteration parameters of $AAC_i^c$, $\alpha_i^c$ and $\beta_i^c$, are extracted from the $i$th layer of Luminance-Net. Then, $AAC_i^c$ is applied to the low-light image input, $I_o$, iteratively, as shown in Figure~\ref{fig:iter}, to get the enhanced result $I_e$.

In Stage-II, we enhance the luminance for the noise-free low-light image and the corresponding image with artificially added Gaussian noise by fixed Luminance-Net. The enhanced noise is more apparent and easier to extract than the noise in the original images. The noise in low-light images is primarily Gaussian~\cite{wei2021physics}, but it becomes unpredictable after enhancement with $AAC_i^c$. We employ DnCNN\cite{zhang2017beyond} architecture as Denoising-Net to estimate it. Furthermore, to more accurately simulate the noise profile in low-light images, we introduce simulate noise larger in the darker image regions as
\begin{equation}
  Noise = (1 - I^c) \otimes N(0,\sigma^c),
  \label{eq:noise}
\end{equation}
where $N(0,\sigma^c)$ is Gaussian distribution map. Because green light is captured using two channels in most camera sensors, green light is generally less noisy than red and blue light.
Therefore, during training, we randomly draw $\sigma\in [0.005, 0.01]$ in every iteration, and set $\sigma^g = \frac{1}{2}\sigma^r = \frac{1}{2}\sigma^b = \sigma$.

\subsection{Loss Function of Stage-I}
We propose a novel and self-reference loss function consisting of four independent components within the estimation factor, which can reflect the optical properties of the real world.
The loss function does not depend on ground truth labels and can guide the training in an unsupervised way.

{\bf Local Color Loss.} The first component of the loss function stems from an assumption based on Retinex theory that the reflectance map is invariant to illumination. We thus enforce the reflectance map of an enhanced image to be close to that of the original image. We propose a local color loss $L_{lcol}$ to preserve the color distribution during the enhancement process:
\begin{equation}
  L_{lcol} = \sum\limits_{c \in \lbrace r, g, b \rbrace}\Vert E{^c_o} - E{^c_e} \Vert_2^2,
  \label{eq:lcol}
\end{equation}
where $E{^c_o}$ and $E{^c_e}$ denote the pixel-wise estimation factor of the original low-light image and enhanced image, respectively. The local color loss function is designed to keep the color distance between two images on the reflectance map normalization plane.

{\bf Global Color Loss.} To avoid over-saturation, we utilize gray-world color constancy hypothesis~\cite{buchsbaum1980spatial} in the design of the global color loss $L_{gcol}$:
\begin{equation}
  L_{gcol} = \sum\limits_{c \in \lbrace r, g, b \rbrace} \Big( A{^c_e}-\frac{1}{3} \Big)^2.
  \label{eq:gcol}
\end{equation}
with
\begin{equation}
    A{^c} = \Big(\sum\limits_{n=1}^N I^c_{(n)}\Big) / \Big(\sum\limits_{n=1}^N \sum\limits_{c' \in \lbrace r, g, b \rbrace} I^{c'}_{(n)}\Big) .
\end{equation}
Here, $A{^c}$ is the average estimation factor value of $c\in \lbrace r, g, b \rbrace$ channel in the whole image.
And, $N$ is the number of pixels in an image. $A{^c_e}$ is $A{^c}$ for the enhanced image.

{\bf Luminance Loss.} We assume that the image sensor is equally sensitive to each RGB channel and has the same light conversion capability. We propose the luminance loss $L_{lum}$, defined as
\begin{equation}
  L_{lum} = \big\Vert H - \sum\limits_{c \in \lbrace r, g, b \rbrace} I{^c_e}\big\Vert_2^2,
  \label{eq:lum}
\end{equation}
with
\begin{equation}
  H = 3\cdot y\cdot\Big(1-\sum\limits_{c \in \lbrace r, g, b \rbrace } \big\Vert E{^c_o} - \frac{1}{3} \big\Vert_2\Big).
  \label{eq:H}
\end{equation}
$H$ represents the expected pixel-wise luminance level, and $y$ is the expected pixel luminance level with reference to the center point of the normalization plane, $[\frac{1}{3}, \frac{1}{3}, \frac{1}{3}] $.
The closer the estimation factor of a pixel is to the center point, the higher the brightness, while the expected luminance decreases when the estimation moves away from the center. We choose $y = 0.8$ in the implementation.

{\bf Curve Smoothness Loss.} To preserve the spatial structure and consistency of the image and avoid undesirable spatial artifacts, we propose a loss function enforcing the adjustment curve smoothness in the spatial dimension. More precisely, we want to minimize the total variation of the $\alpha$ and $\beta$ components used in the definition of AAC. This loss component is defined as follows
\begin{equation}
  L_{smo}(\zeta)=\frac{1}{N}\sum\limits_{c\in\lbrace r, g, b \rbrace}\Vert\bigtriangledown\zeta^c\Vert_2^2, \zeta\in\{\alpha,\beta\}.
  \label{eq:smooth}
\end{equation}

{\bf Total Loss Function.} The total loss function combines together all the sub-loss functions defined above
\begin{equation}
  L_1 = w_{lcol}\cdot L_{lcol} + w_{gcol} \cdot L_{gcol} + w_{lum} \cdot L_{lum}+ w_{\zeta}\cdot L_{smo}(\zeta).
  \label{eq:total}
\end{equation}

We set the weights corresponding to different components of the loss function as $w_{lcol}=1000$, $w_{gcol}=1500$, $w_{lum}=5$, $w_{\alpha}=1000$ and $w_{\beta}=5000$. We set $w_{\beta}$ to be larger than $w_{\alpha}$, as the function $\beta$ should not change rapidly and break the local monotonicity of the adjustment curve. The weight factors $w$ are hyper-parameters determining the final behavior of our enhancement method. Increasing $ w_{gcol}$ and decreasing $w_{lcol}$ limit the image saturation. Increasing $w_{lum}$ leads to improved image brightness, but it can also introduce additional noise in the dark image areas.

\begin{figure*}[htb]
  \centering

  \begin{subfigure}{0.19\linewidth}
    \includegraphics[width=0.62\linewidth]{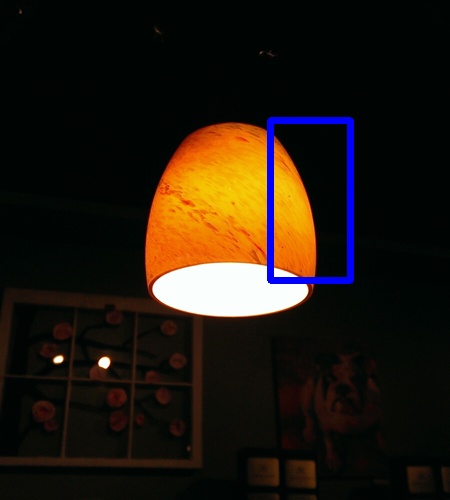} 
    \includegraphics[width=0.345\linewidth]{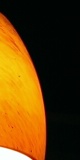} 
    \caption{Input}
    \label{fig:input}
    \end{subfigure}
  \hfill
    \begin{subfigure}{0.19\linewidth}
    \includegraphics[width=0.62\linewidth]{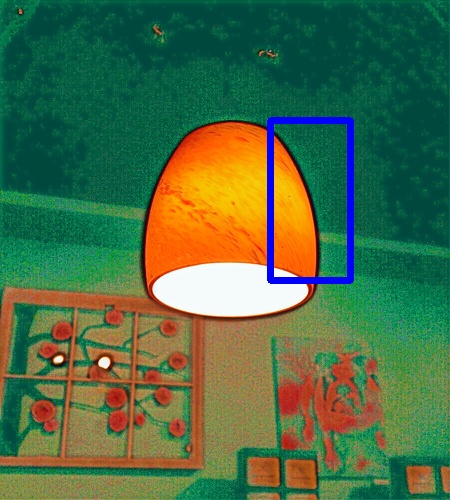}
    \includegraphics[width=0.345\linewidth]{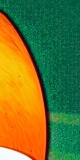}
    \caption{Retinex-Net}
    \label{fig:retinex}
    \end{subfigure}
  \hfill
    \begin{subfigure}{0.19\linewidth}
    \includegraphics[width=0.62\linewidth]{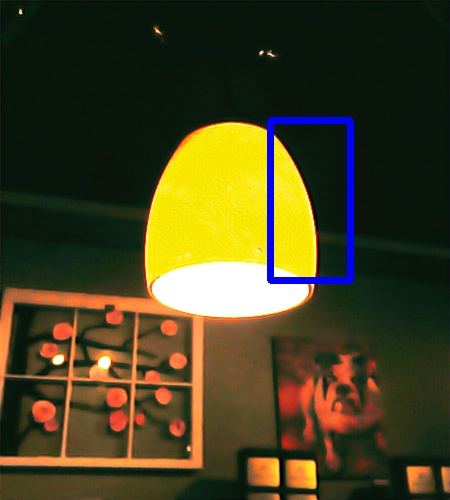} 
    \includegraphics[width=0.345\linewidth]{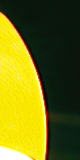} 
    \caption{RUAS}
    \label{fig:RUAS}
    \end{subfigure}
  \hfill
    \begin{subfigure}{0.19\linewidth}
    \includegraphics[width=0.62\linewidth]{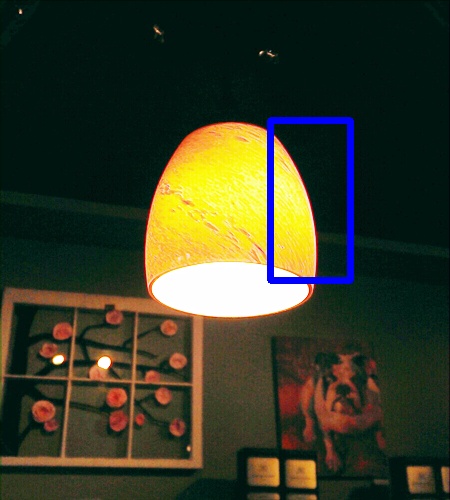} 
    \includegraphics[width=0.345\linewidth]{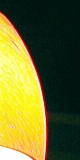} 
    \caption{SCI}
    \label{fig:sci}
    \end{subfigure}
  \hfill
    \begin{subfigure}{0.19\linewidth}
    \includegraphics[width=0.62\linewidth]{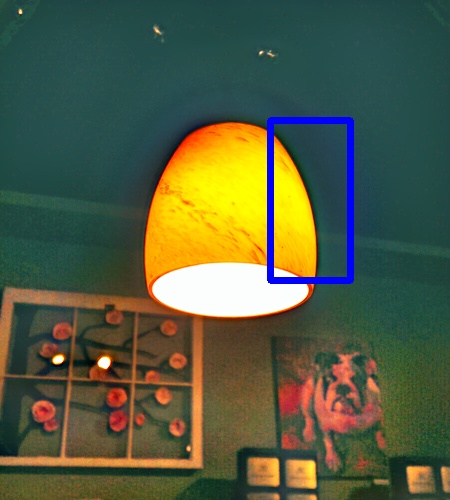} 
    \includegraphics[width=0.345\linewidth]{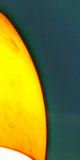} 
    \caption{EnGAN}
    \label{fig:EnGAN}
    \end{subfigure}
  \hfill
    \begin{subfigure}{0.19\linewidth}
    \includegraphics[width=0.62\linewidth]{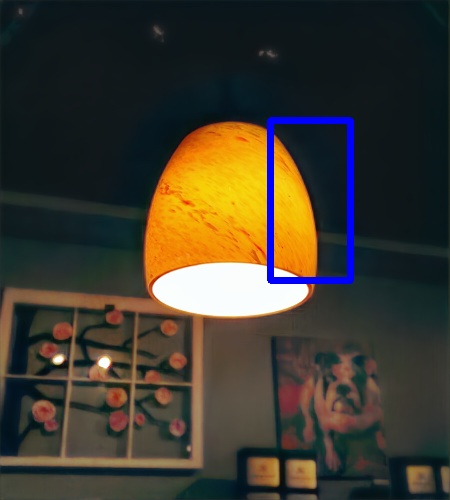} 
    \includegraphics[width=0.345\linewidth]{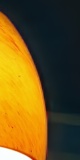} 
    \caption{KinD}
    \label{fig:kind}
    \end{subfigure}
  \hfill
    \begin{subfigure}{0.19\linewidth}
    \includegraphics[width=0.62\linewidth]{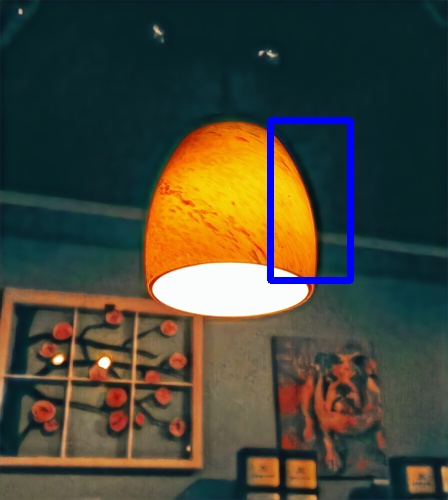} 
    \includegraphics[width=0.345\linewidth]{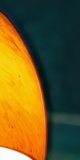} 
    \caption{KinD++}
    \label{fig:kindpp}
    \end{subfigure}
  \hfill
  \begin{subfigure}{0.19\linewidth}
    \includegraphics[width=0.62\linewidth]{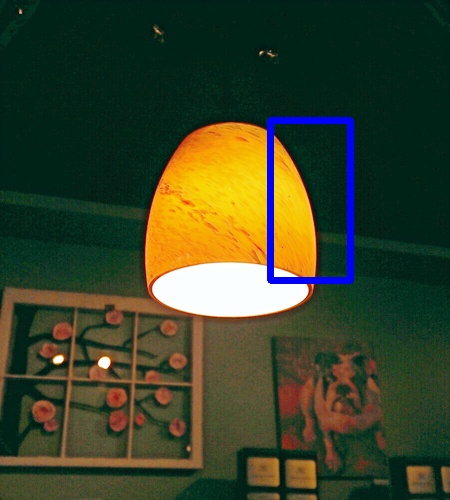} 
    \includegraphics[width=0.345\linewidth]{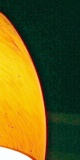} 
    \caption{ZeroDCE}
    \label{fig:DCE}
    \end{subfigure}
  \hfill
    \begin{subfigure}{0.19\linewidth}
    \includegraphics[width=0.62\linewidth]{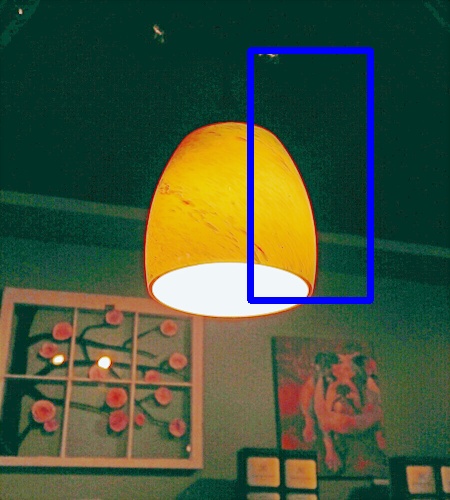} 
    \includegraphics[width=0.345\linewidth]{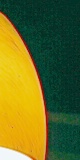} 
    \caption{Ours$^\ast$}
    \label{fig:ours1}
    \end{subfigure}
  \hfill
    \begin{subfigure}{0.19\linewidth}
    \includegraphics[width=0.62\linewidth]{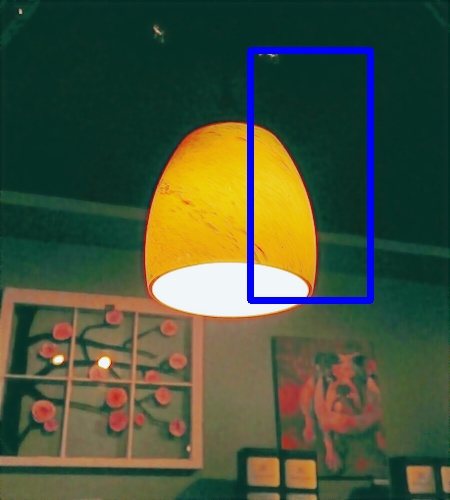} 
    \includegraphics[width=0.345\linewidth]{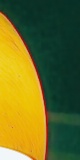} 
    \caption{Ours}
    \label{fig:ours}
    \end{subfigure}

  \caption{Visual Comparison on LIME. The blue box zooms in the complex light and dark junction of the input image. Image of Ours$^\ast$ is the output from Stage-I.}
  \label{fig:comparison}
\end{figure*}

\subsection{Loss Function of Stage-II}

Stage-II aims to estimate enhanced noise previously hidden in the dark images. In Figure~\ref{fig:frame},
$I_d$ is the enhanced result after denoising a simulated noisy image, and $I_e$ is obtained by enhancing the noise-free image. We assume $I_e$ approximates the optimal solution, set as a pseudo ground truth for the denoising task.

To this end, we utilize SSIM~\cite{wang2004image} to preserve the image structure. We also found that keeping the gradient of both images aligned achieves superior denoising performance compared to using PSNR. Moreover, we add $ \Vert \bigtriangledown I_d\Vert_2^2$ to accelerate the convergence of denoising training process. The total loss can be formulated as

\begin{equation}
  L_2 = -w_s\cdot SSIM(I_e, I_d) + w_g\cdot \Vert \bigtriangledown I_e - \bigtriangledown I_d\Vert_2^2 + \Vert \bigtriangledown I_d\Vert_2^2.
  \label{eq:total_2}
\end{equation}

We set $w_s = 10$ and $w_g = 40$, because $\Vert \bigtriangledown I_e - \bigtriangledown I_d\Vert_2^2$ is the core of denoising for Stage-II.

\section{Results}
In this section, we present implementation details and an ablation study investigating the significance of the main components of our algorithm. We also show a qualitative comparison between images enhanced using our method and other existing algorithms. Next, we present exhaustive quantitative comparisons with existing state-of-the-art algorithms applied to three real-world low-light image datasets. 

\subsection{Implementation Details}
We train Self-DACE in both Stage-I and Stage-II on 2002 images of SCIE~Part1~\cite{cai2018learning} dataset, similarly as proposed in~\cite{guo2020zero}. SCIE dataset contains noise-free images taken in various complex lighting conditions, and each set includes multi-level exposed images. This diversity helps Self-DACE learn to enhance both over- and underexposed images. To reduce the training time, we resize all training images to $512\times 512$.

We trained and tested Self-DACE on NVIDIA RTX 3080-Ti GPU using ADAM optimizer and PyTorch framework. We set the batch size to $8$, weight decay to $1e^{-4}$, and use a constant leaning rate of $1e^{-4}$. In Stage-I, Luminance-Net is lightweight and the training converges after 200 epochs, corresponding to about 30 minutes. It only takes around 0.1 millisecond to process a $512\times 512$ RGB image In Stage-II, Denoising-Net takes about half an hour to converge after around 50 epochs. It takes approximately only 0.01 second to enhance a $512\times 512$ RGB image.

\begin{table}[t]
\centering
\begin{tabular}{cccc} 
\toprule
Stage                     & Ablation Item & PSNR↑  & SSIM↑  \\ 
\midrule
\multirow{4}{*}{Stage-I}  & w/o $L_{lum}$  & 5.09  & 0.20  \\
                          & w/o $L_{lcol}$ & 13.50 & 0.61  \\
                          & w/o $L_{gcol}$ & 18.42 & 0.65  \\
                          & w/o $L_{smo}$ & 14.31 & 0.52  \\ 
\midrule
\multirow{3}{*}{Stage-II} & w/o SSIM  & 18.99 & 0.76  \\
                          & w/o $\Vert \bigtriangledown I_e - \bigtriangledown I_d\Vert_2^2$      & 19.01 & 0.75  \\
                          & w/o $\Vert \bigtriangledown I_d\Vert_2^2$        & 19.11 & 0.78  \\
\bottomrule
\end{tabular}
\vspace{1em}
\caption{Ablation study of the effect of each loss component.}
\label{tab:loss_ab}
\end{table}

\begin{table}[t]
\centering
\begin{tabular}{cccc} 
\toprule
Network                     & Ablation Item  & PSNR↑  & SSIM↑  \\ 
\midrule
\multirow{4}{*}{Luminance-Net}  & 5-layer  & 18.21 & 0.50  \\
                          & 6-layer  & 18.48 & 0.58  \\
                          & 8-layer  & 18.53 & 0.58  \\
                          & 9-layer  & 18.56 & 0.59  \\ 
\midrule
\multirow{4}{*}{Denoising-Net} & 18-layer & 18.99 & 0.74  \\
                          & 19-layer & 19.05 & 0.75  \\
                          & 21-layer & 19.18 & 0.78  \\
                          & 22-layer & 19.19 & 0.78  \\
\bottomrule
\end{tabular}
\vspace{1em}
\caption{Ablation study of the effect of layer structures.}
\label{tab:net_ab}
\vspace{-1em}
\end{table}

\begin{table*}[t]
\centering
\begin{tabular}{ccccccccc>{\columncolor[gray]{.9} }c>{\columncolor[gray]{.9} }c} 
\toprule
\multirow{2}{*}[-0.5ex]{Datasets}    & \multirow{2}{*}[-0.5ex]{Metrics} & \multicolumn{3}{c}{Supervised Learning}                                         & \multicolumn{6}{c}{Unsupervised Learning}                                                                                        \\ 
\cmidrule(lr){3-5} \cmidrule(lr){6-11}
                             &                          & Retinex-Net & KinD                            & KinD++                           & EnGAN                            & ZeroDCE & RUAS  & SCI   & Ours$^\ast$              & Ours                              \\ 
\midrule
\multirow{4}{*}{LOL\_test}   & PSNR↑                    & 16.77       & 17.65                           & 17.75                            & 17.48                            & 14.86   & 16.40 & 14.78 & \textcolor{blue}{\textbf{18.50}} & \textcolor{red}{\textbf{19.15}}   \\
                             & SSIM↑                    & 0.42        & \textcolor{blue}{\textbf{0.77}} & \textcolor{blue}{\textbf{0.77}}  & 0.65                             & 0.56    & 0.50  & 0.52  & 0.58                             & \textcolor{red}{\textbf{0.78}}    \\
                             & LPIPS↓                   & 0.474                 & \textcolor{red}{\textbf{0.175}} & 0.198  & 0.322 & 0.335                           & 0.270 & 0.339                            & 0.332                           & \textcolor{blue}{\textbf{0.183}}  \\
                             & CIEDE↓                   & 16.47       & 16.63                           & \textcolor{red}{\textbf{14.74}}  & 17.91                            & 25.47   & 22.90 & 30.41 & 15.85                            & \textcolor{blue}{\textbf{15.33}}  \\ 
\midrule
\multirow{4}{*}{LSRW}        & PSNR↑                    & 15.51       & 16.41                           & 16.08                            & 17.08 & 15.86   & 14.27 & 15.24 & \textcolor{blue}{\textbf{17.10}}                            & \textcolor{red}{\textbf{17.24}}   \\
                             & SSIM↑                    & 0.37        & \textcolor{blue}{\textbf{0.48}} & 0.40                             & 0.47                             & 0.45    & 0.47  & 0.42  & 0.46                             & \textcolor{red}{\textbf{0.52}}    \\
                             & LPIPS↓                   & 0.427                 & 0.337                           & 0.371  & 0.327 & \textcolor{red}{\textbf{0.318}} & 0.465 & \textcolor{blue}{\textbf{0.322}} & 0.346                           & 0.368  \\
                             & CIEDE↓                   & 19.88       & 18.26                           & 17.83                            & \textcolor{red}{\textbf{16.26}}  & 21.25   & 23.29 & 25.71 & 17.60                            & \textcolor{blue}{\textbf{17.43}}  \\ 
\midrule
\multirow{4}{*}{SICE\_Part2} & PSNR↑                    & 19.90       & 19.60                           & 19.92                            & 18.47                            & 16.78   & 14.53 & 15.74 & \textcolor{red}{\textbf{20.90}}  & \textcolor{blue}{\textbf{20.67}}  \\
                             & SSIM↑                    & 0.73        & 0.77                            & 0.77                             & 0.77                             & 0.74    & 0.69  & 0.71  & \textcolor{red}{\textbf{0.82}}   & \textcolor{blue}{\textbf{0.79}}   \\
                             & LPIPS↓                   & 0.422                 & 0.398                           & 0.399  & 0.404 & 0.397                           & 0.526 & \textcolor{blue}{\textbf{0.392}} & \textcolor{red}{\textbf{0.321}} & 0.394  \\
                             & CIEDE↓                   & 11.59       & 13.29                           & 11.68                            & 14.44                            & 18.69   & 19.73 & 22.43 & \textcolor{red}{\textbf{10.73}}  & \textcolor{blue}{\textbf{10.88}}  \\
\bottomrule
\end{tabular}
  \caption{Quantitative comparisons in terms of four full-reference image quality metrics including PSNR(dB), SSIM, LPIPS and CIEDE2000 on the LOL$\_$test, LSRW and SCIE$\_$Part2 datasets. The best result is in \textcolor{red}{\textbf{red}}, the sub-optimal result is in \textcolor{blue}{\textbf{blue}}. Ours$^\ast$ is the result only from Stage-I.}
  \label{tab:comparison}
\end{table*}

\begin{table}[t]
\centering
\begin{tabular}{cccc} 
\toprule
                    & Method      & Parameters(M)↓                   & FLOPs(G)↓                          \\ 
\midrule
\multirow{3}{*}{SL} & Retinex-Net & 0.555                            & 587.47                             \\
                    & KinD        & 8.160                             & 574.954                            \\
                    & KinD++      & 8.275                            & 12238.03                           \\ 
\midrule
\multirow{6}{*}{UL} & EnGAN       & 8.637                            & 273.24                             \\
                    & ZeroDCE     & 0.079                            & 84.99                              \\
                    & RUAS        & \textcolor{red}{\textbf{0.003}}  & \textcolor{red}{\textbf{3.528}}    \\
                    & SCI         & \textcolor{blue}{\textbf{0.011}} & 188.873                            \\ 

                    & \cellcolor[gray]{.9}Ours$^\ast$        & \cellcolor[gray]{.9}0.068                            & \cellcolor[gray]{.9}\textcolor{blue}{\textbf{73.716}}  \\

                    & \cellcolor[gray]{.9}Ours        & \cellcolor[gray]{.9}0.699                            & \cellcolor[gray]{.9}754.272                            \\
\bottomrule
\end{tabular}
\vspace{1em}
\caption{Comparisons of computational complexity in terms of number of trainable parameters and FLOPs. The best result is in \textcolor{red}{\textbf{red}}, the sub-optimal result is in \textcolor{blue}{\textbf{blue}}. UL, SL mean unsupervised and supervised learning, respectively. Ours$^\ast$ is the model of Stage-I, and Ours is the model including Stage-I and Stage-II.}
\label{tab:size}
\vspace{-1em}
\end{table}

\subsection{Ablation Study}
We investigate the contributions of all
loss function components and the effect of network structures to the final performance of the trained network for 2 stages separately.
We include an example of the results tested in LoL\_test dataset with evaluation metrics, PSNR, SSIM, referring to the ground truth. 
As shown is Table\ref{tab:loss_ab},
$L_{lum}$ added to preserve the image luminosity has the most noticeable impact on the result of Self-DACE.
Removing $L_{lcol}$ leads to substantial color distortion, with the intensity of each color channel increasing and making the image visibly overexposed. It causes a significant drop in the metric values consequently.
The global color loss $L_{gcol}$ corrects some of the color cast and leads to visually better results. It manifests as a slight yellow tint for almost all images and leads to a slight decrease in the metrics.
$L_{smo}$ connects the network and the pixel-wise adjustment curve and enforces the local monotonicity on the enhancement curves. Without applying $L_{smo}$, the adjustment curve can lose its monotonicity, resulting in color distortion and increased noise. 
For Stage-II, each loss component only has a small impact on the results, while the results can't get the best with one of them absent.

According to Table\ref{tab:net_ab},
less layers both in Luminance-Net and Denoising-Net results worse performance and make the network hard to train, while a larger network brings only a minor benefit.

\subsection{Comparison with the State-of-the-art}
To demonstrate the effectiveness and generalizability of our algorithm, compare it to eight other leading methods on three different real-world low-light image datasets across three metrics.

We perform the analysis on the LOL$\_$test dataset~\cite{wei2018deep}, LSRW dataset~\cite{hai2021r2rnet}, and SCIE$\_$Part2 dataset~\cite{cai2018learning}. In the SCIE$\_$Part2 dataset, we choose the first 125 subsets, with the first and second dark levels for each subset, totaling 250 images and we resize all images to $512 \times 512$ in the SCIE$\_$Part2, for some images in SCIE is too large for some methods with big networks to process them.

As there is a lack of specialized and effective non-reference metrics for low-light enhancement, we prioritize full-reference metrics with predefined desired outputs. Our selection comprises four commonly used metrics for assessing low-light image enhancement: PSNR, SSIM~\cite{wang2004image}, LPIPS\cite{zhang2018unreasonable} and CIEDE2000\cite{sharma2005ciede2000}. 
LPIPS is a famous metric that serves human visual perception and is commonly used in low-level image quality assessment. 
LPIPS is a deep learning-based and provides three different pre-trained networks. We adopt the AlexNet-based model for its outstanding performance and moderate size. Moreover, CIEDE2000 is a metric developed by ISO/CIE that focuses on measuring color differences.

We compare our method with existing SOTA algorithms, including
Retinex-Net~\cite{wei2018deep}, KinD~\cite{zhang2019kindling}, KinD++~\cite{zhang2021beyond}, EnGAN~\cite{jiang2021enlightengan}, ZeroDCE~\cite{guo2020zero}, RUSA~\cite{liu2021retinex}, and SCI~\cite{ma2022toward}. 

Table~\ref{tab:comparison} contains a summary of a quantitative comparison of Self-DACE with other low-light enhancement algorithms listed above. Our method consistently outperforms all algorithms in almost all metrics on all considered datasets. And the results from Stage-I without denoising also achieve the the best or second best results on multiple datasets.

As shown in Table~\ref{tab:size}, which provides the computational complexity comparison of trainable parameters and FLOPs when those methods are applied to a 1200$\times$900 RGB image, Self-DACE is relative smaller than KinD, KinD++ and EnGAN. Stage-I of Self-DACE has the advantage of being lightweight and computationally fast while obtained sub-optimal metric values.

In Figure~\ref{fig:comparison}, we present a visual comparison of an image sampled from the LIME dataset~\cite{Lime2017}. When applied to a low-light image in complex lighting conditions, Retinex-Net visibly distorts the colors. EnGAN introduces helo-like artifacts, while KinD++ and ZeroDCE add artifacts on the edges between bright and dark regions of the input image. It can be observed near the boundary of the hanging lamp as a thick black line surrounding it.  SCI, RUAS and KinD cannot illuminate the dark parts of the picture where bright areas exist, among which especially RUAS adds a too-strong enhancement leading to an impression of an overexposed image and loss of important details and texture characteristics. Comparing to luminance enhancement only, the final result of 2-stage is more smooth and have a better visual effect.
Our method visibly outperforms all algorithms discussed above in removing artifacts and enhancing dark areas while preserving details of brighter parts.


\section{Conclusion}
This article proposed an unsupervised and self-reference 2-stage low-light enhancement method - Self-Reference Deep Adaptive Curve Estimation for Low-Light Image Enhancement. 
The first stage is based on Retinex theory and simplified physical model and can be realized by a lightweight neural network Luminance-Net. 
We trained Luminance-Net to output smooth image enhancement curves, which are then iteratively applied to enhance the images. 
The Second stage uses Denoising-Net to estimate the noise hidden in the dark after luminance enhancement.
Extensive experiments show that our algorithm outperforms existing methods visually and quantitatively on several benchmarks. 

\bibliographystyle{unsrt}  
\bibliography{references} 
\newpage

\begin{appendices}
\renewcommand{\thefigure}{A1}
\begin{figure*}[b]

  \centering
  \begin{subfigure}{0.19\linewidth}
    \includegraphics[width=1\linewidth]{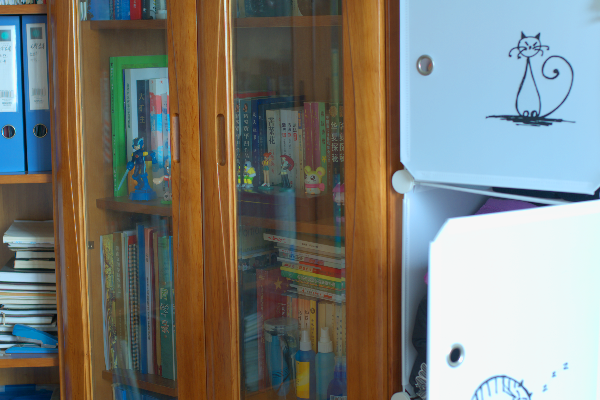} 
    \caption{GT}
    \label{fig:gt}
    \end{subfigure}
  \hfill
    \begin{subfigure}{0.19\linewidth}
    \includegraphics[width=1\linewidth]{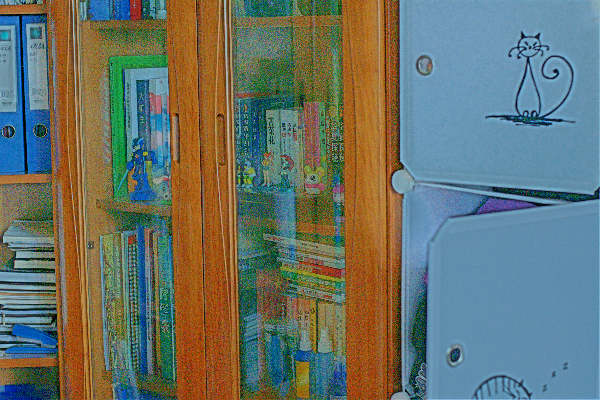} 
    \caption{Retinex-Net}
    \label{fig:Retinex_face}
    \end{subfigure}
  \hfill
    \begin{subfigure}{0.19\linewidth}
    \includegraphics[width=1\linewidth]{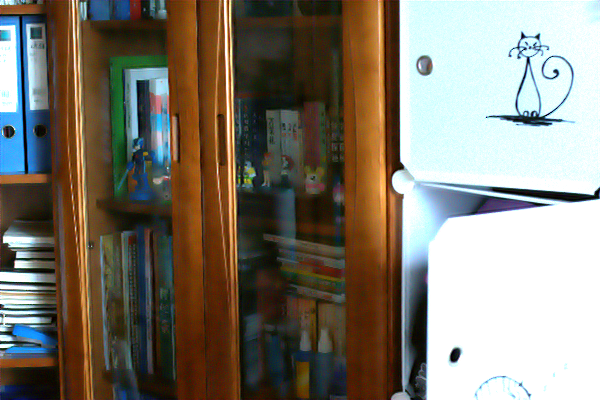} 
    \caption{RUAS}
    \label{fig:RUAS_face}
    \end{subfigure}
  \hfill
  \begin{subfigure}{0.19\linewidth}
    \includegraphics[width=1\linewidth]{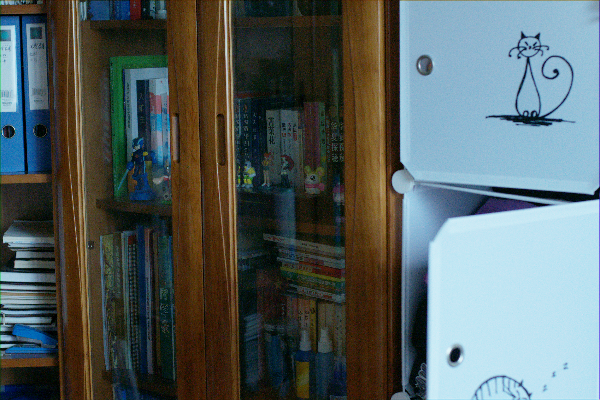} 
    \caption{SCI}
    \label{fig:sci_face}
    \end{subfigure}
  \hfill
  \begin{subfigure}{0.19\linewidth}
    \includegraphics[width=1\linewidth]{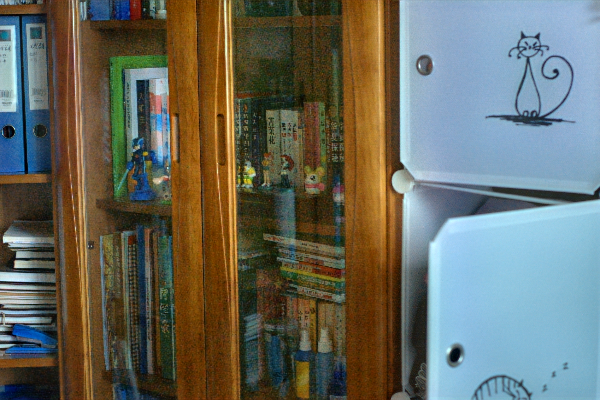} 
    \caption{EnlightenGAN}
    \label{fig:EnlightenGAN_face}
    \end{subfigure}
  \hfill
  \begin{subfigure}{0.19\linewidth}
    \includegraphics[width=1\linewidth]{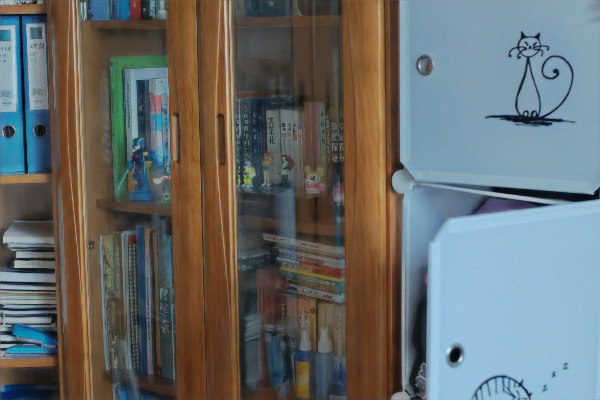} 
    \caption{KinD}
    \label{fig:KinD_face}
    \end{subfigure}
  \hfill
    \begin{subfigure}{0.19\linewidth}
    \includegraphics[width=1\linewidth]{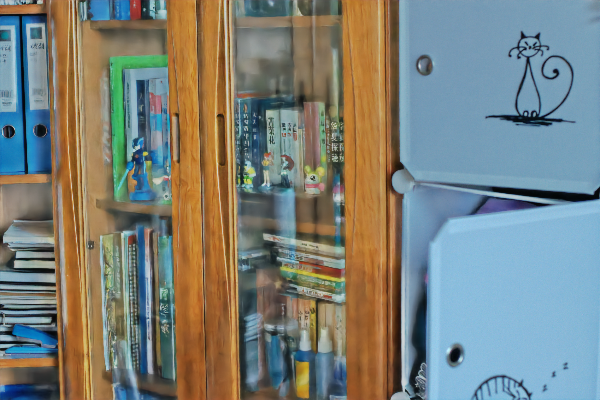} 
    \caption{KinD++}
    \label{fig:KinD++_face}
    \end{subfigure}
  \hfill
    \begin{subfigure}{0.19\linewidth}
    \includegraphics[width=1\linewidth]{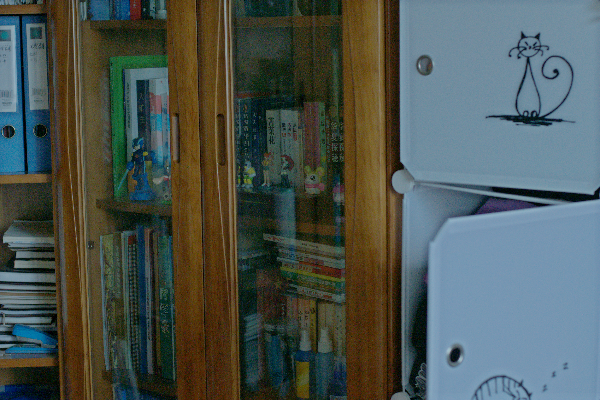} 
    \caption{ZeroDCE}
    \label{fig:dce_face}
    \end{subfigure}
  \hfill
  \begin{subfigure}{0.19\linewidth}
    \includegraphics[width=1\linewidth]{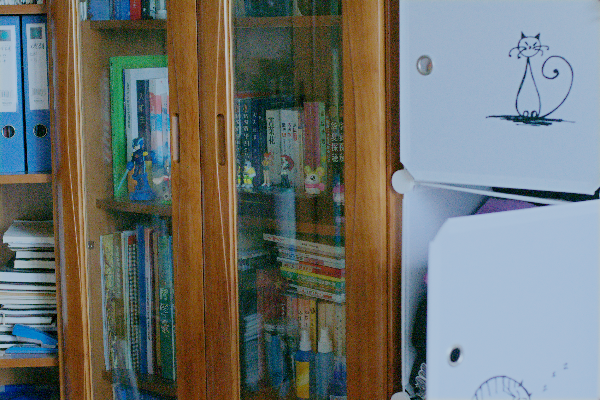} 
    \caption{Ours$^\ast$}
    \label{fig:DRBN_face}
    \end{subfigure}
  \hfill
    \begin{subfigure}{0.19\linewidth}
    \includegraphics[width=1\linewidth]{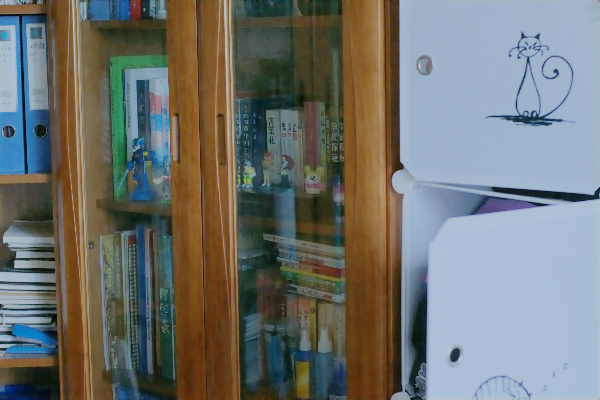} 
    \caption{Ours}
    \label{fig:Ours_face}
    \end{subfigure}
  \caption{Visual comparison for the 1.png of LoL$\_$test Dataset.}
  \label{fig:LOL1}
\end{figure*}

\renewcommand{\thefigure}{A2}
\begin{figure*}[b]

  \centering
  \begin{subfigure}{0.19\linewidth}
    \includegraphics[width=1\linewidth]{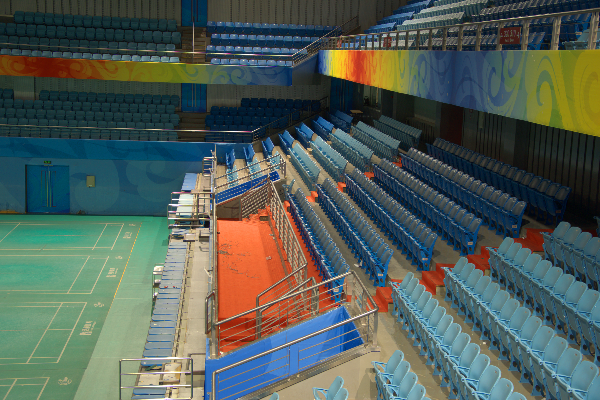} 
    \caption{GT}
    \label{fig:gt}
    \end{subfigure}
  \hfill
    \begin{subfigure}{0.19\linewidth}
    \includegraphics[width=1\linewidth]{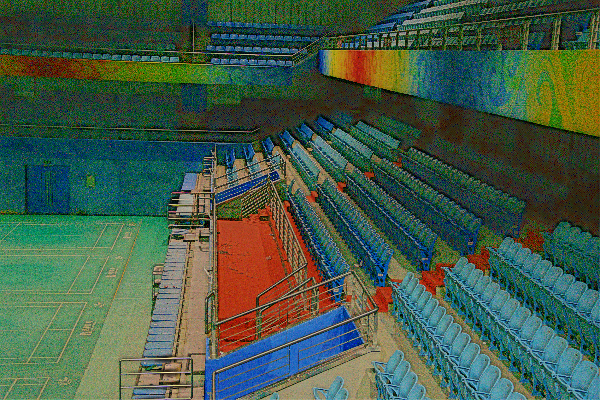} 
    \caption{Retinex-Net}
    \label{fig:Retinex_face}
    \end{subfigure}
  \hfill
    \begin{subfigure}{0.19\linewidth}
    \includegraphics[width=1\linewidth]{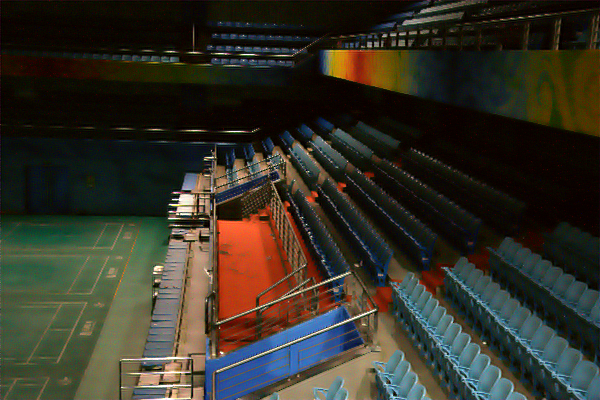} 
    \caption{RUAS}
    \label{fig:RUAS_face}
    \end{subfigure}
  \hfill
  \begin{subfigure}{0.19\linewidth}
    \includegraphics[width=1\linewidth]{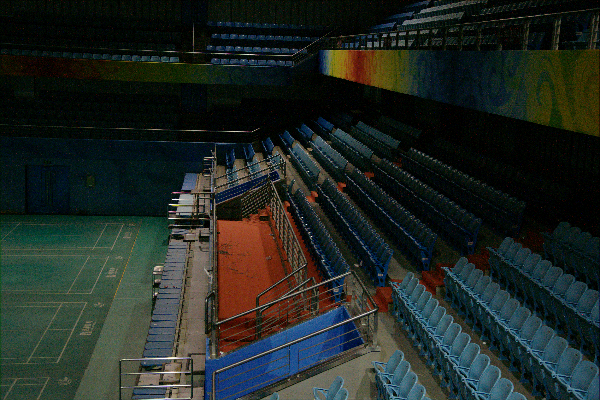} 
    \caption{SCI}
    \label{fig:sci_face}
    \end{subfigure}
  \hfill
  \begin{subfigure}{0.19\linewidth}
    \includegraphics[width=1\linewidth]{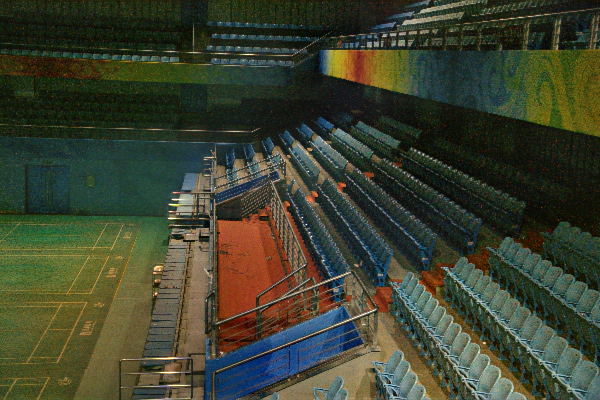} 
    \caption{EnlightenGAN}
    \label{fig:EnlightenGAN_face}
    \end{subfigure}
  \begin{subfigure}{0.19\linewidth}
    \includegraphics[width=1\linewidth]{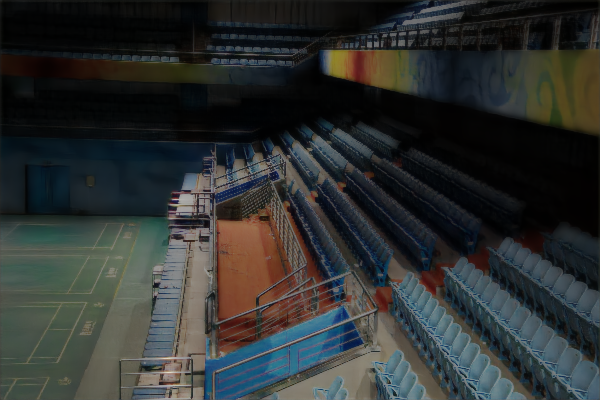} 
    \caption{KinD}
    \label{fig:KinD_face}
    \end{subfigure}
  \hfill
    \begin{subfigure}{0.19\linewidth}
    \includegraphics[width=1\linewidth]{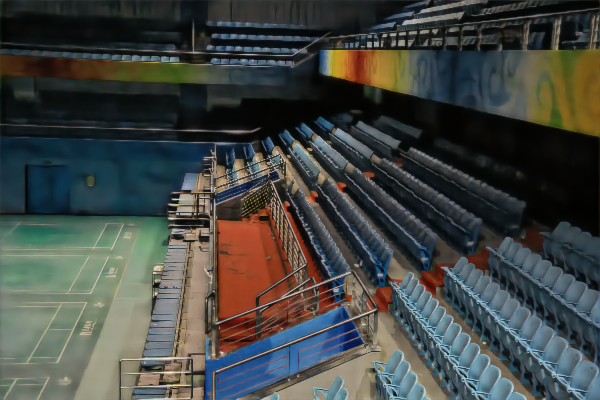} 
    \caption{KinD++}
    \label{fig:KinD++_face}
    \end{subfigure}
  \hfill
    \begin{subfigure}{0.19\linewidth}
    \includegraphics[width=1\linewidth]{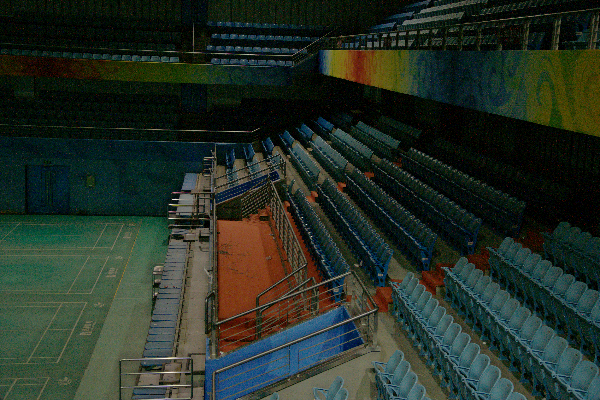} 
    \caption{ZeroDCE}
    \label{fig:dce_face}
    \end{subfigure}
  \hfill
  \begin{subfigure}{0.19\linewidth}
    \includegraphics[width=1\linewidth]{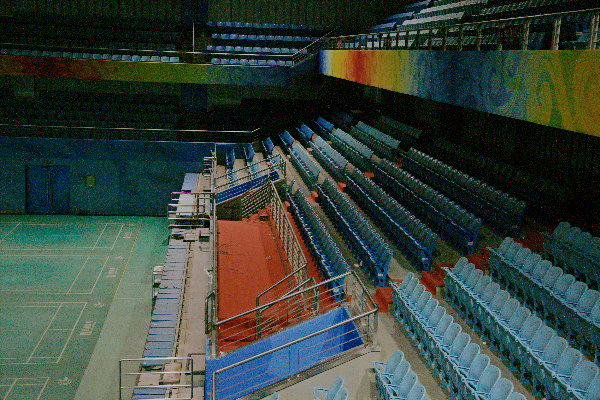} 
    \caption{Ours$^\ast$}
    \label{fig:DRBN_face}
    \end{subfigure}
  \hfill
  \hfill
    \begin{subfigure}{0.19\linewidth}
    \includegraphics[width=1\linewidth]{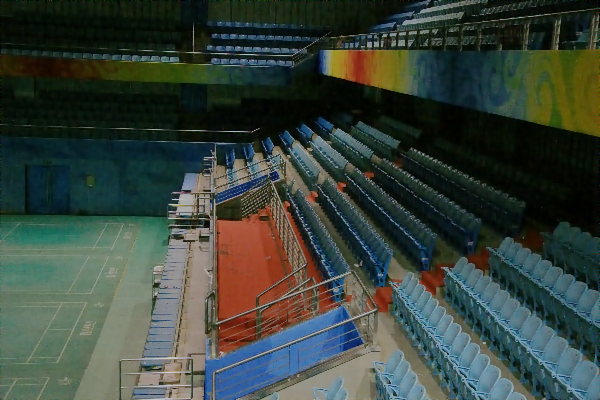} 
    \caption{Ours}
    \label{fig:ours}
    \end{subfigure}
  \caption{Visual comparison for the 778.png of LoL$\_$test Dataset.}
  \label{fig:LOL5}
\end{figure*}
\renewcommand{\thefigure}{A3}
\begin{figure*}[b]

  \centering
  \begin{subfigure}{0.19\linewidth}
    \includegraphics[width=1\linewidth]{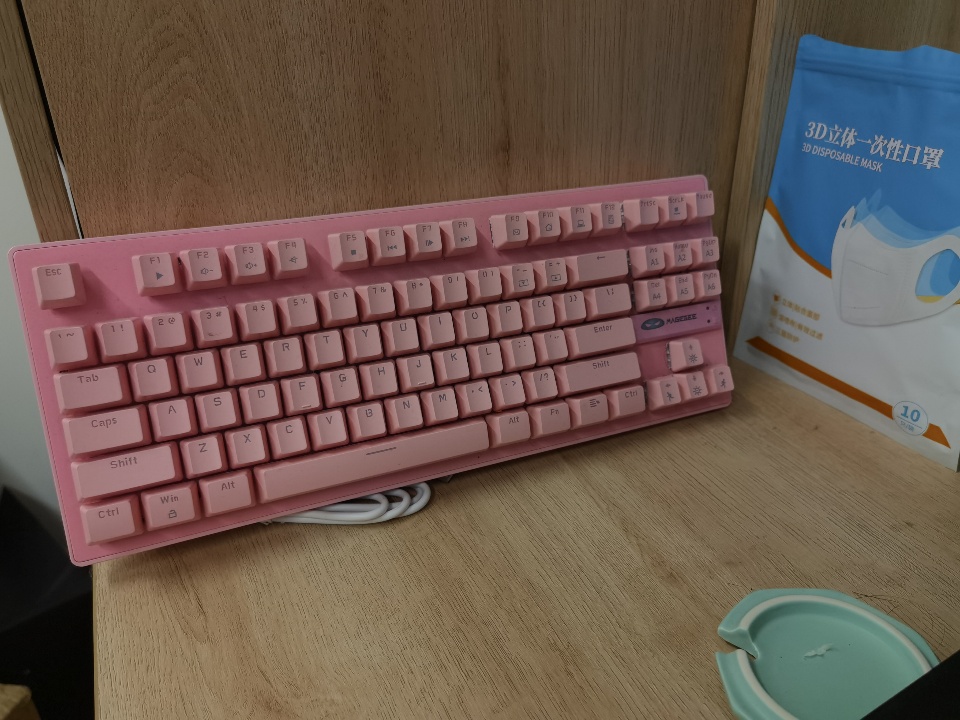} 
    \caption{GT}
    \label{fig:gt}
    \end{subfigure}
  \hfill
    \begin{subfigure}{0.19\linewidth}
    \includegraphics[width=1\linewidth]{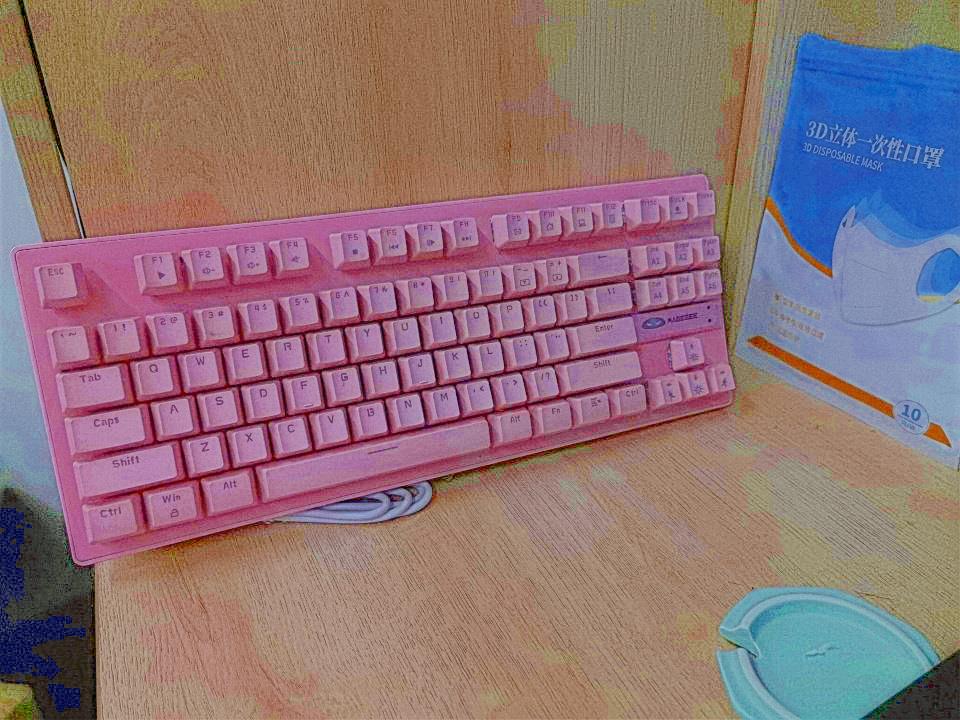} 
    \caption{Retinex-Net}
    \label{fig:Retinex_face}
    \end{subfigure}
  \hfill
    \begin{subfigure}{0.19\linewidth}
    \includegraphics[width=1\linewidth]{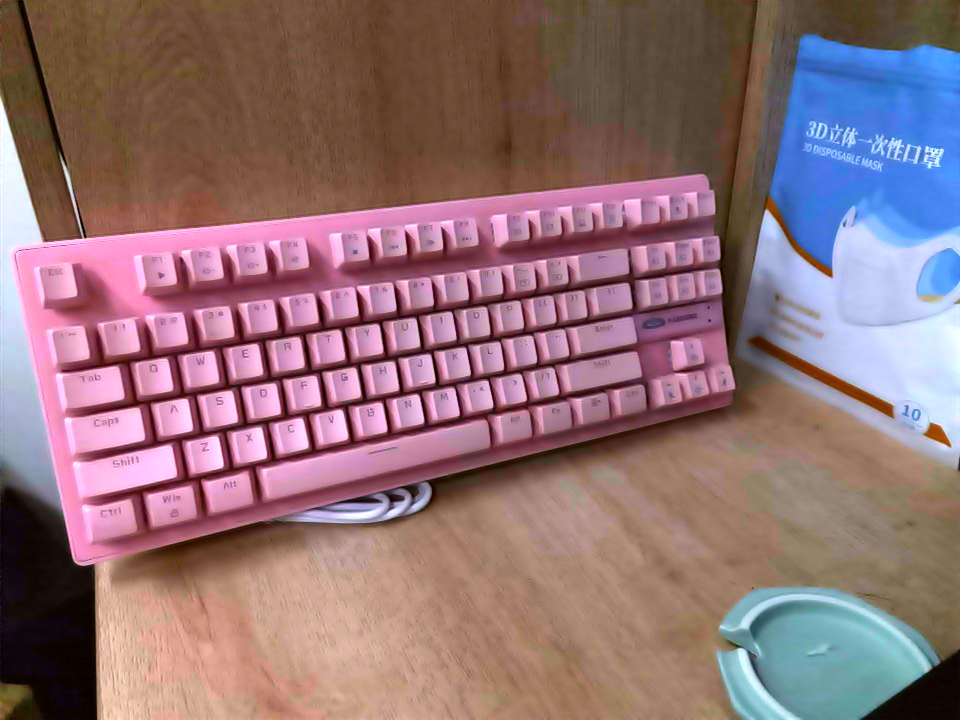} 
    \caption{RUAS}
    \label{fig:RUAS_face}
    \end{subfigure}
  \hfill
  \begin{subfigure}{0.19\linewidth}
    \includegraphics[width=1\linewidth]{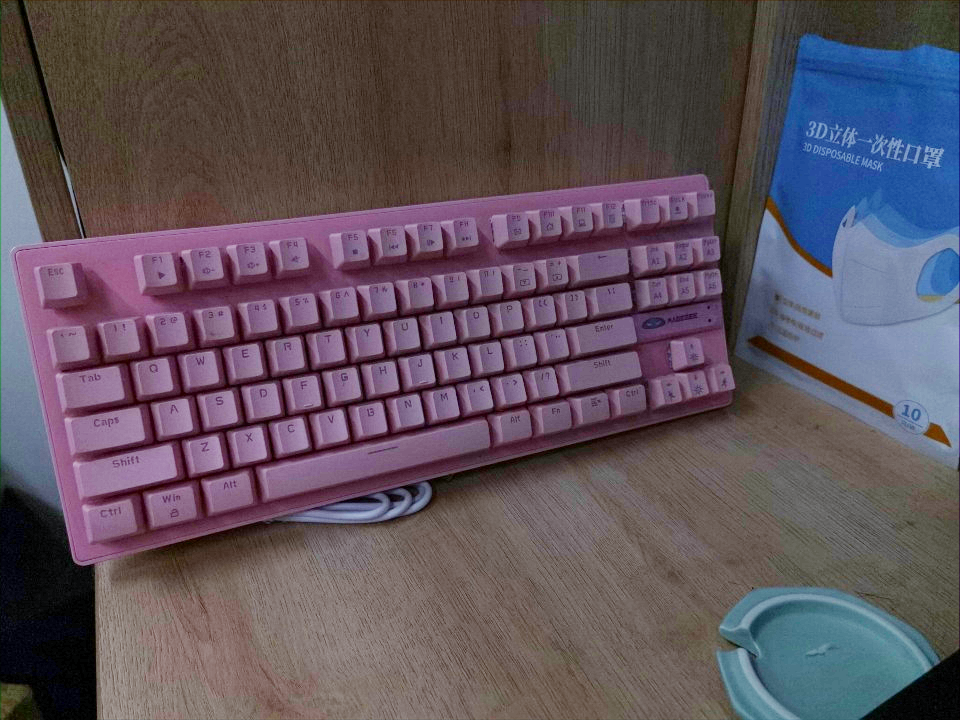} 
    \caption{SCI}
    \label{fig:sci_face}
    \end{subfigure}
  \hfill
  \begin{subfigure}{0.19\linewidth}
    \includegraphics[width=1\linewidth]{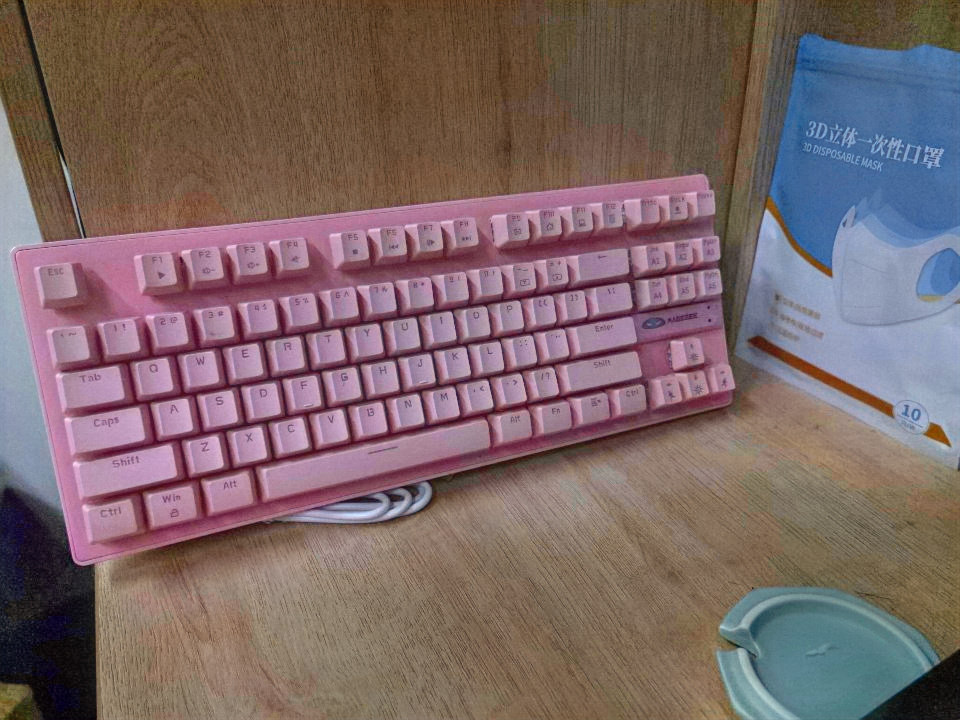} 
    \caption{EnlightenGAN}
    \label{fig:EnlightenGAN_face}
    \end{subfigure}
  \hfill
  \begin{subfigure}{0.19\linewidth}
    \includegraphics[width=1\linewidth]{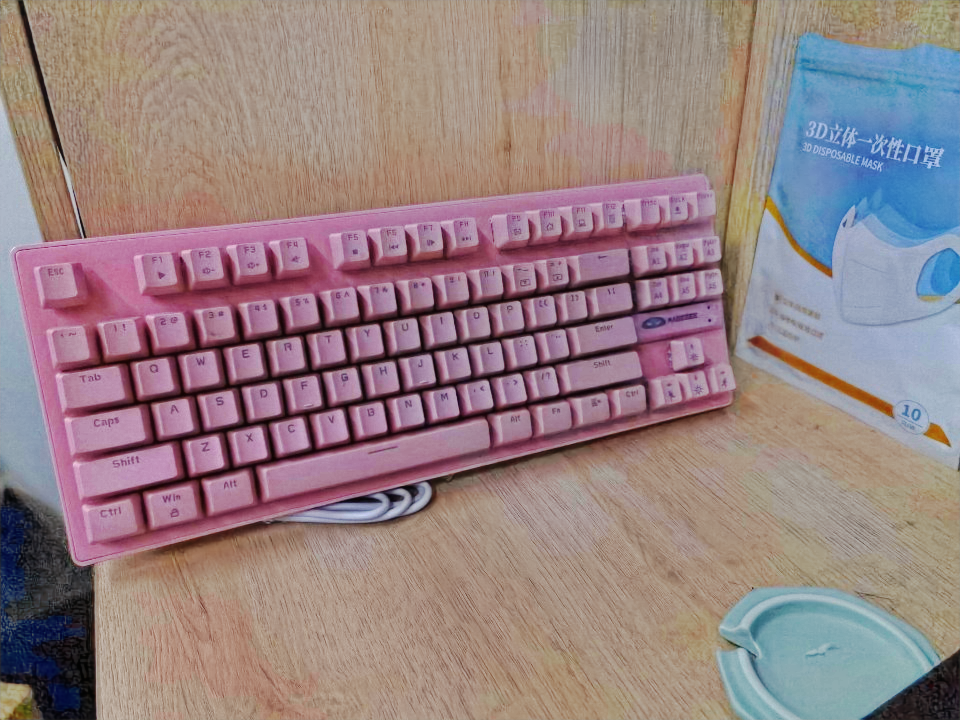} 
    \caption{KinD}
    \label{fig:KinD_face}
    \end{subfigure}
  \hfill
    \begin{subfigure}{0.19\linewidth}
    \includegraphics[width=1\linewidth]{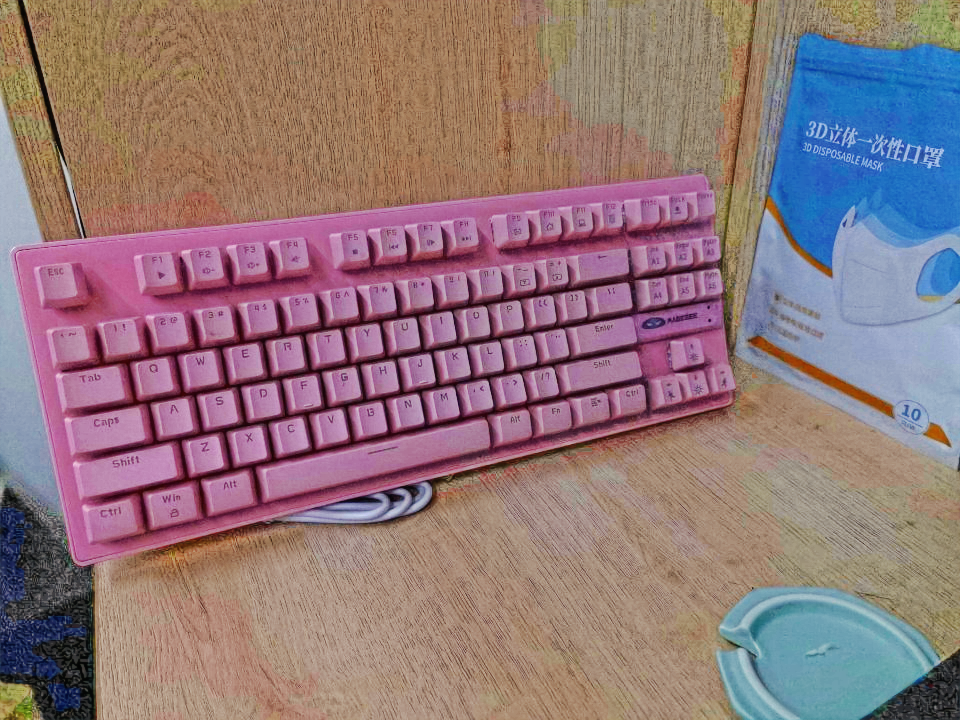} 
    \caption{KinD++}
    \label{fig:KinD++_face}
    \end{subfigure}
  \hfill
    \begin{subfigure}{0.19\linewidth}
    \includegraphics[width=1\linewidth]{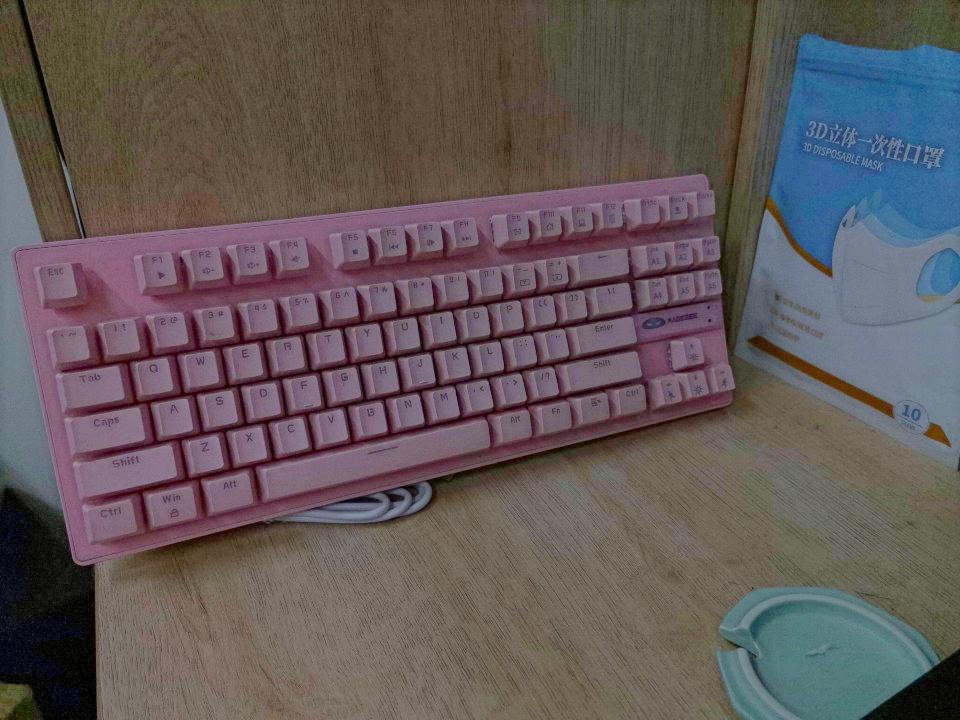} 
    \caption{ZeroDCE}
    \label{fig:dce_face}
    \end{subfigure}
  \hfill
  \begin{subfigure}{0.19\linewidth}
    \includegraphics[width=1\linewidth]{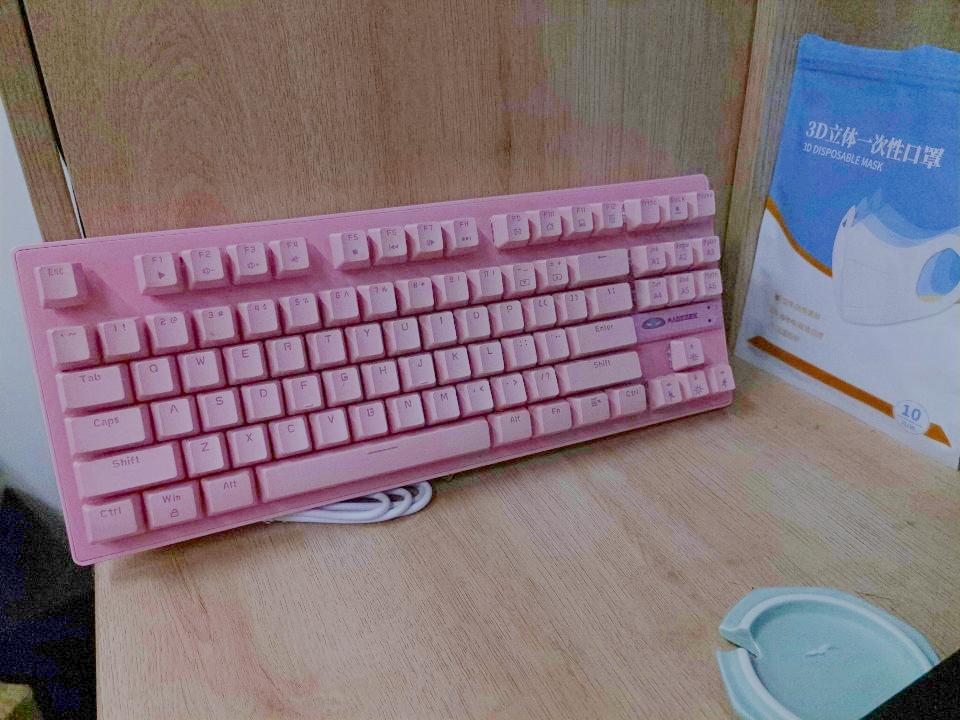} 
    \caption{Ours$^\ast$}
    \label{fig:DRBN_face}
    \end{subfigure}
  \hfill
    \begin{subfigure}{0.19\linewidth}
    \includegraphics[width=1\linewidth]{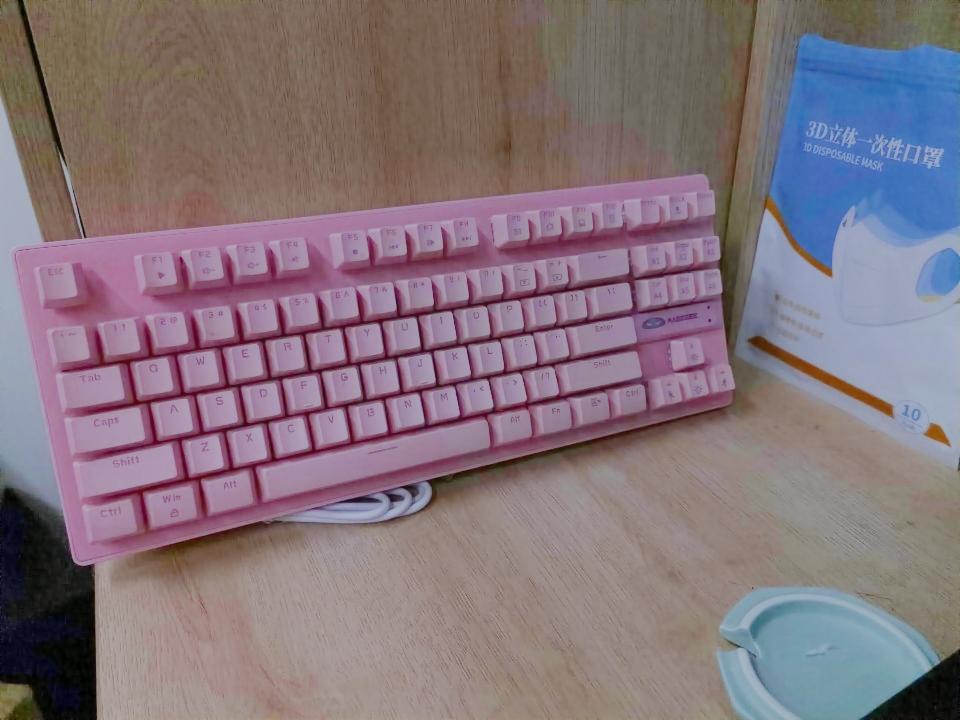} 
    \caption{Ours}
    \label{fig:Ours_face}
    \end{subfigure}
  \caption{Visual comparison for the 2060.jpg of LSRW Dataset.}
  \label{fig:LSRW3}
\end{figure*}
\renewcommand{\thefigure}{A4}
\begin{figure*}[b]
\vspace{-0.3cm}
\setlength{\abovecaptionskip}{0cm} 
\setlength{\belowcaptionskip}{-0.1cm} 
  \centering
  \begin{subfigure}{0.19\linewidth}
    \includegraphics[width=1\linewidth]{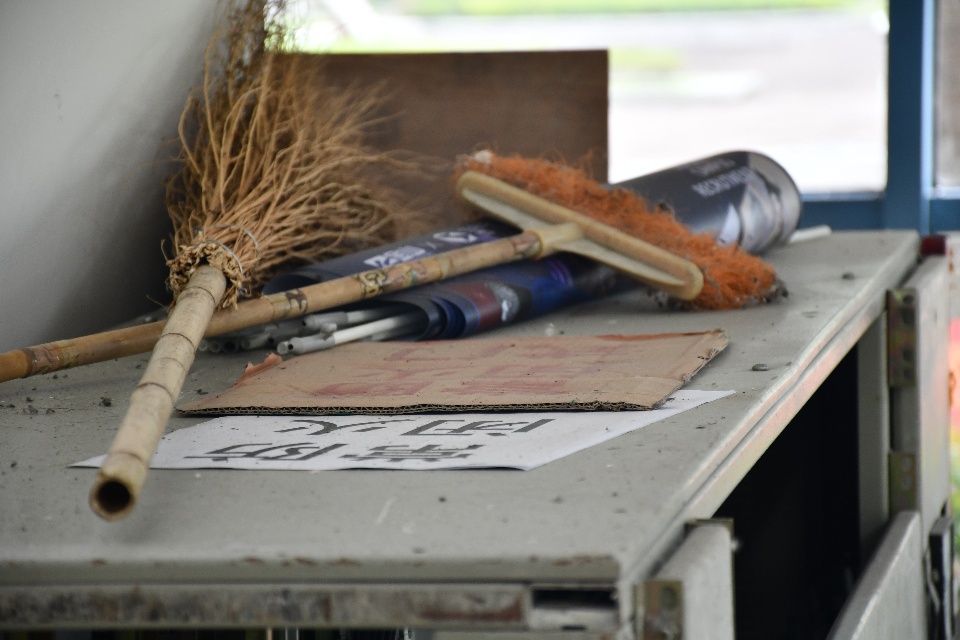} 
    \caption{GT}
    \label{fig:gt}
    \end{subfigure}
  \hfill
    \begin{subfigure}{0.19\linewidth}
    \includegraphics[width=1\linewidth]{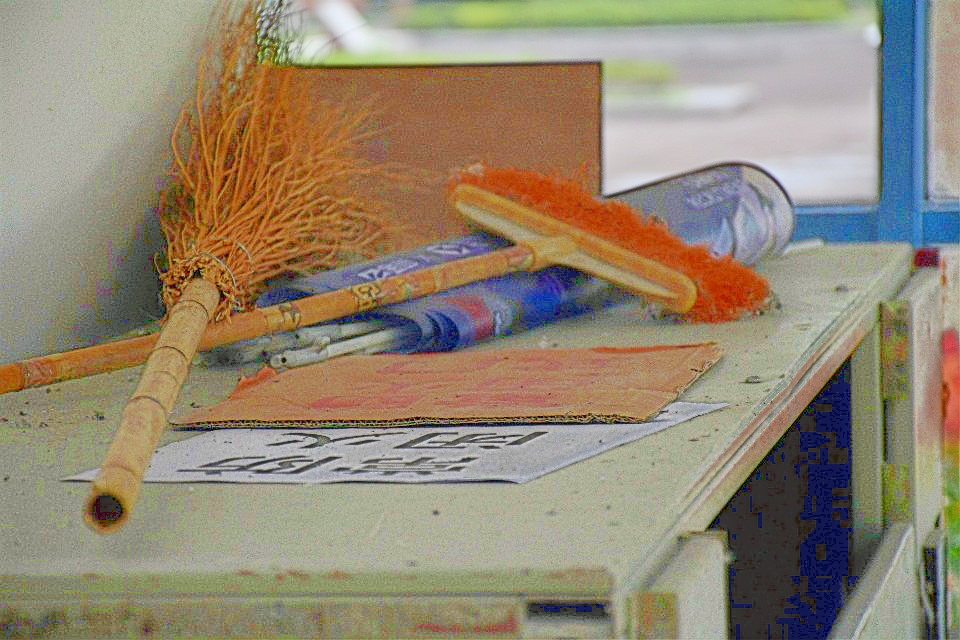} 
    \caption{Retinex-Net}
    \label{fig:Retinex_face}
    \end{subfigure}
  \hfill
    \begin{subfigure}{0.19\linewidth}
    \includegraphics[width=1\linewidth]{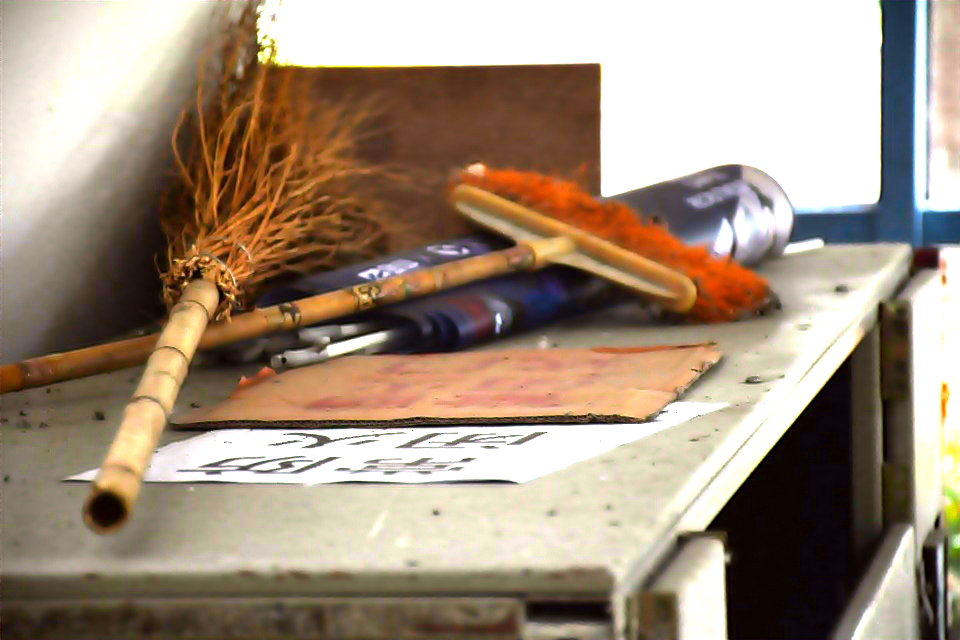} 
    \caption{RUAS}
    \label{fig:RUAS_face}
    \end{subfigure}
  \hfill
  \begin{subfigure}{0.19\linewidth}
    \includegraphics[width=1\linewidth]{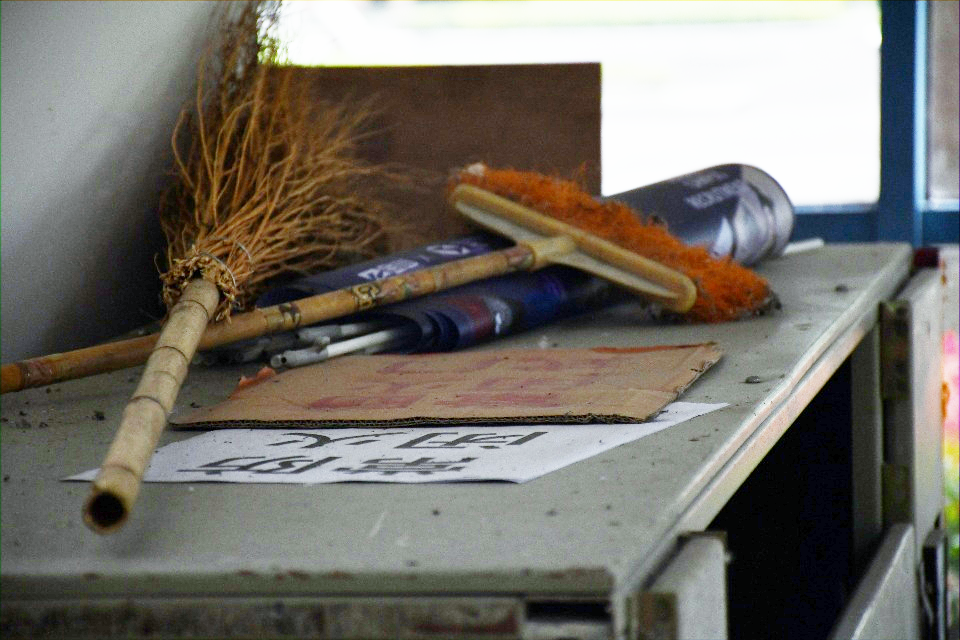} 
    \caption{SCI}
    \label{fig:sci_face}
    \end{subfigure}
  \hfill
  \begin{subfigure}{0.19\linewidth}
    \includegraphics[width=1\linewidth]{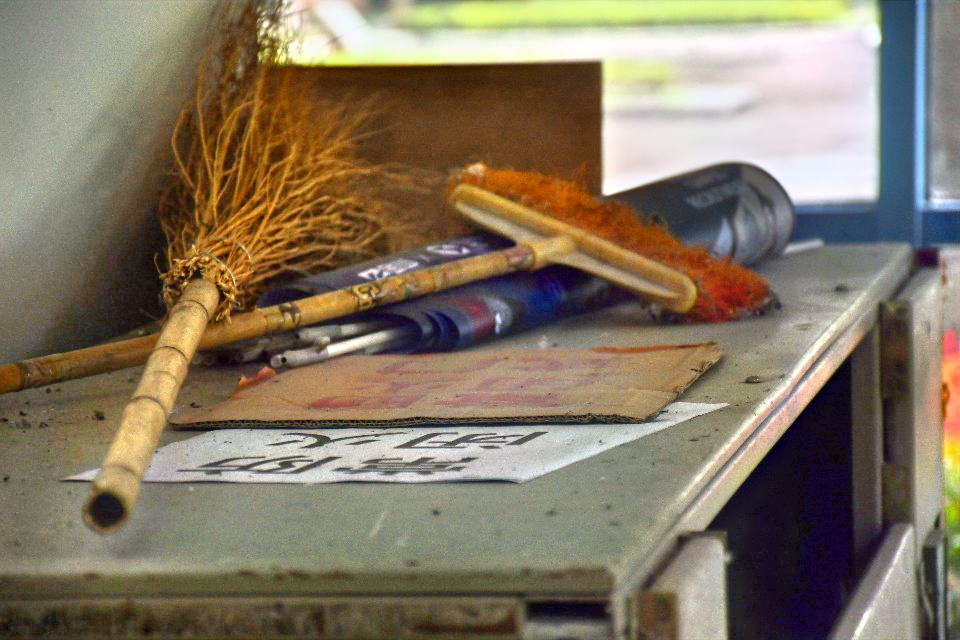} 
    \caption{EnlightenGAN}
    \label{fig:EnlightenGAN_face}
    \end{subfigure}
  \hfill
  \begin{subfigure}{0.19\linewidth}
    \includegraphics[width=1\linewidth]{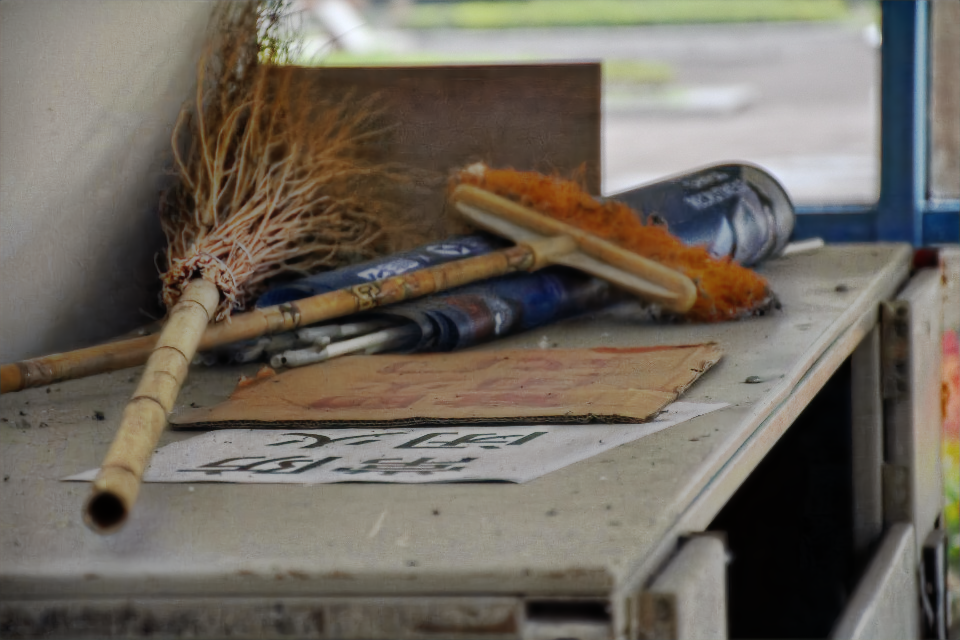} 
    \caption{KinD}
    \label{fig:KinD_face}
    \end{subfigure}
  \hfill
    \begin{subfigure}{0.19\linewidth}
    \includegraphics[width=1\linewidth]{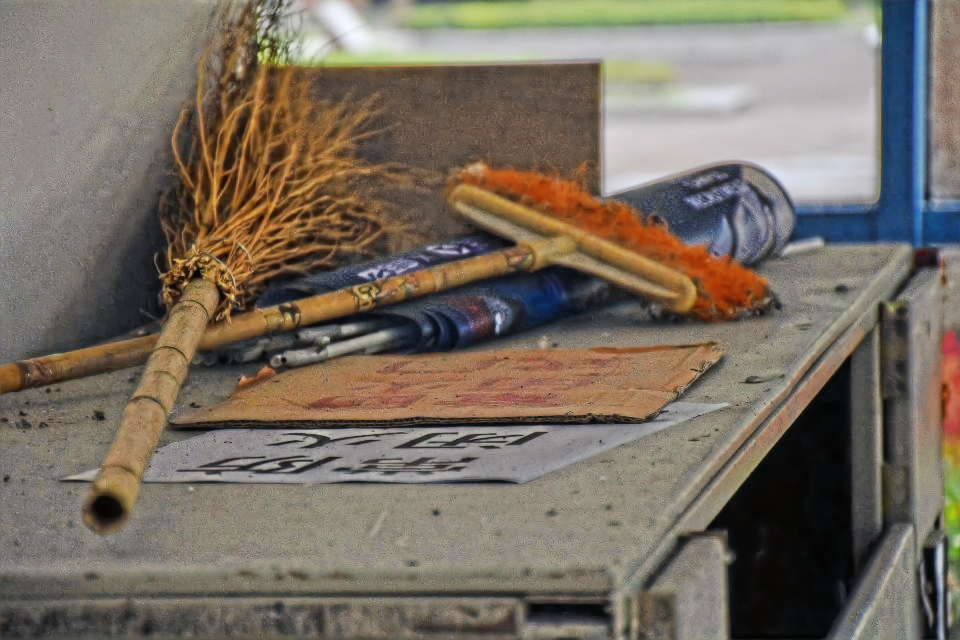} 
    \caption{KinD++}
    \label{fig:KinD++_face}
    \end{subfigure}
  \hfill
    \begin{subfigure}{0.19\linewidth}
    \includegraphics[width=1\linewidth]{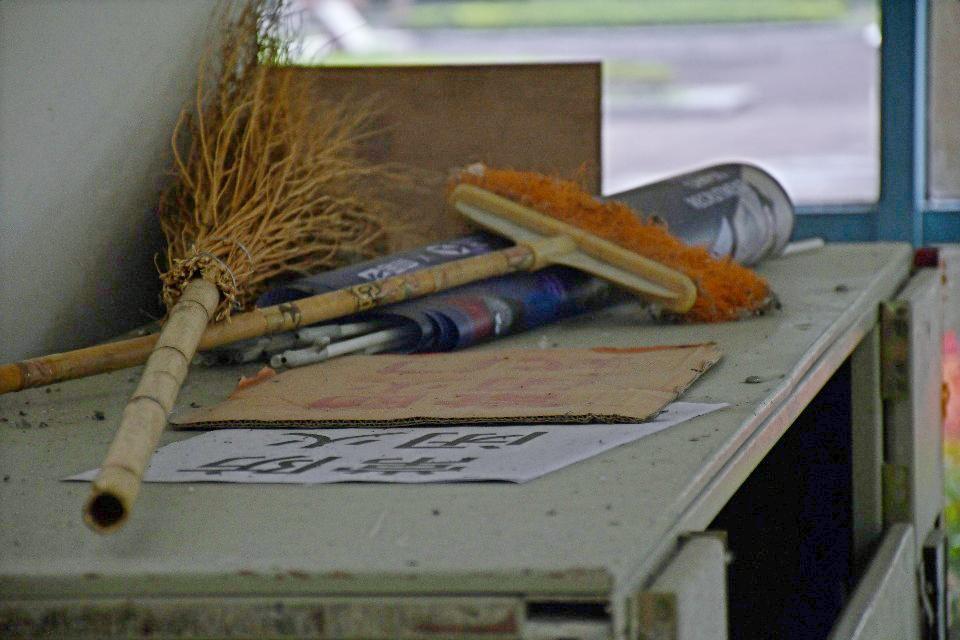} 
    \caption{ZeroDCE}
    \label{fig:dce_face}
    \end{subfigure}
  \hfill
  \begin{subfigure}{0.19\linewidth}
    \includegraphics[width=1\linewidth]{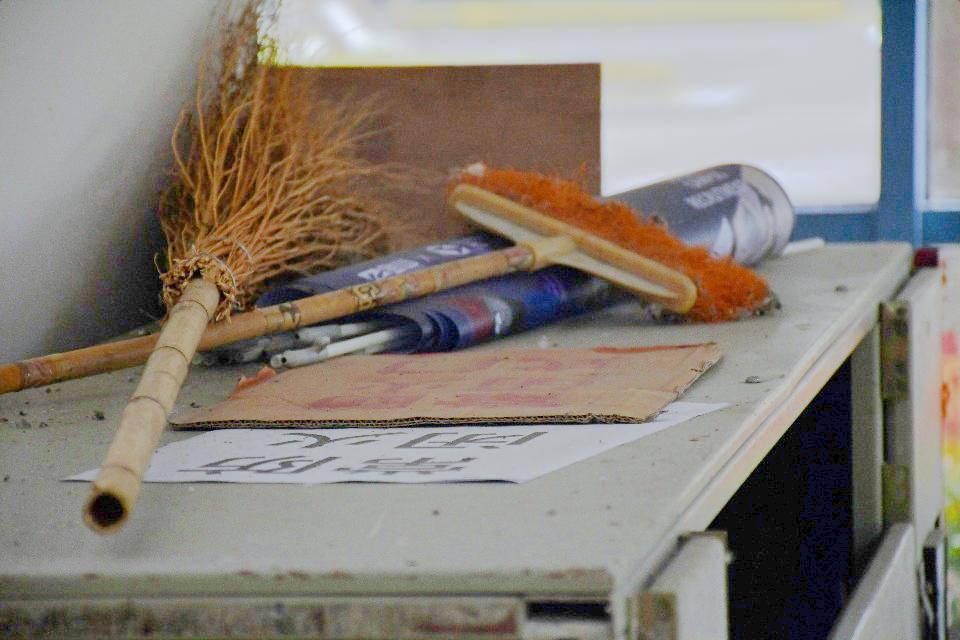} 
    \caption{Ours$^\ast$}
    \label{fig:DRBN_face}
    \end{subfigure}
  \hfill
    \begin{subfigure}{0.19\linewidth}
    \includegraphics[width=1\linewidth]{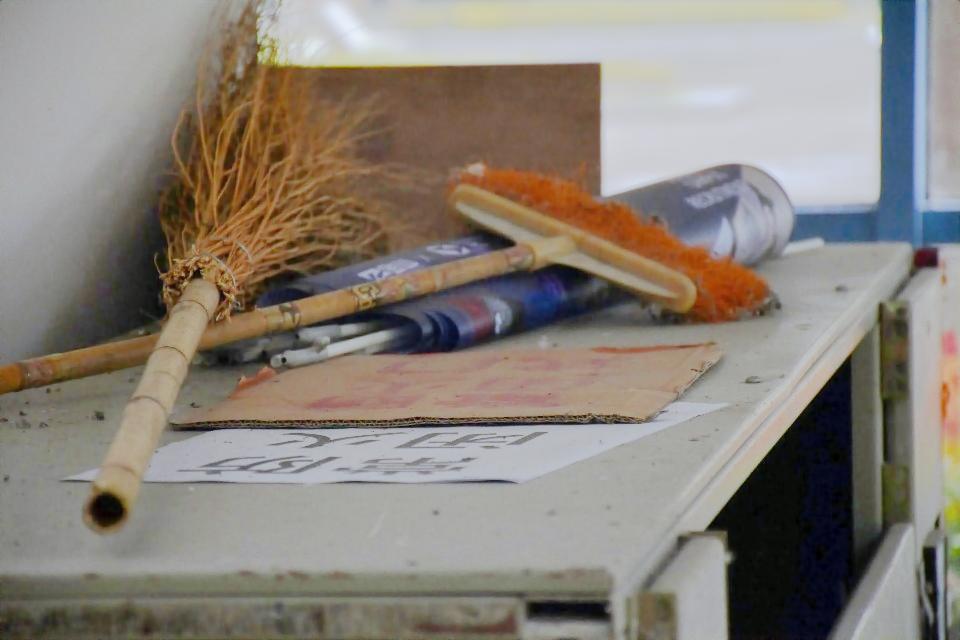} 
    \caption{Ours}
    \label{fig:Ours_face}
    \end{subfigure}
  \caption{Visual comparison for the 3020.jpg of LSRW Dataset.}
  \label{fig:LSRW5}
\end{figure*}
\renewcommand{\thefigure}{A5}
\begin{figure*}[b]
\vspace{-0.3cm}
\setlength{\abovecaptionskip}{0cm} 
\setlength{\belowcaptionskip}{-0.1cm} 
  \centering
  \begin{subfigure}{0.19\linewidth}
    \includegraphics[width=1\linewidth]{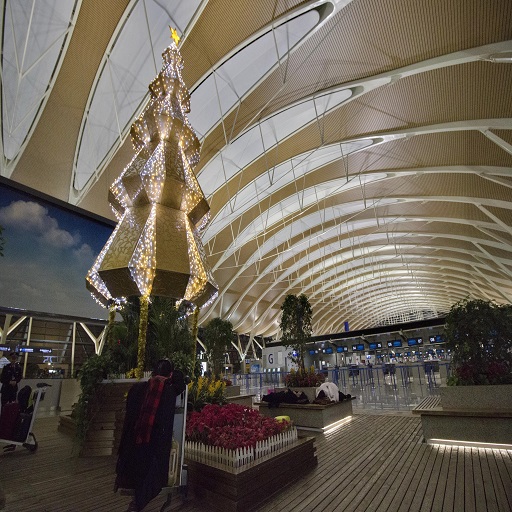} 
    \caption{GT}
    \label{fig:gt}
    \end{subfigure}
  \hfill
    \begin{subfigure}{0.19\linewidth}
    \includegraphics[width=1\linewidth]{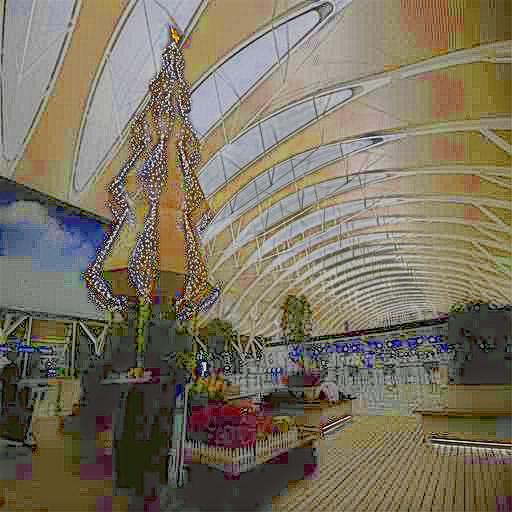} 
    \caption{Retinex-Net}
    \label{fig:Retinex_face}
    \end{subfigure}
  \hfill
    \begin{subfigure}{0.19\linewidth}
    \includegraphics[width=1\linewidth]{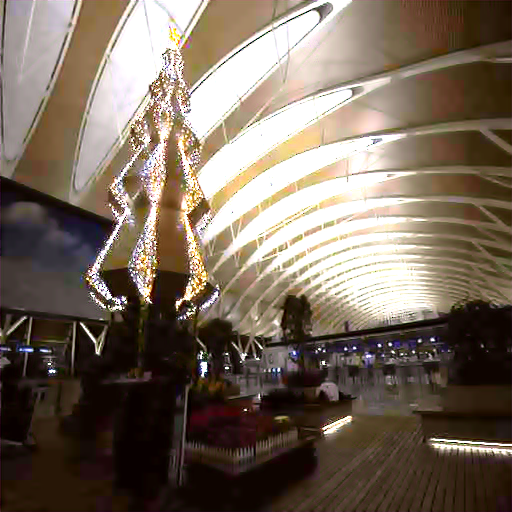} 
    \caption{RUAS}
    \label{fig:RUAS_face}
    \end{subfigure}
  \hfill
  \begin{subfigure}{0.19\linewidth}
    \includegraphics[width=1\linewidth]{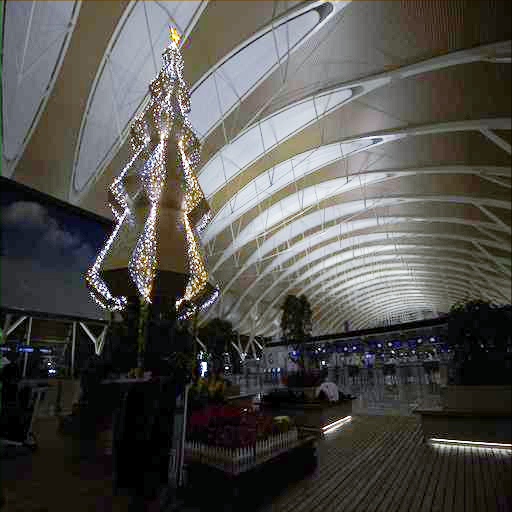} 
    \caption{SCI}
    \label{fig:sci_face}
    \end{subfigure}
  \hfill
  \begin{subfigure}{0.19\linewidth}
    \includegraphics[width=1\linewidth]{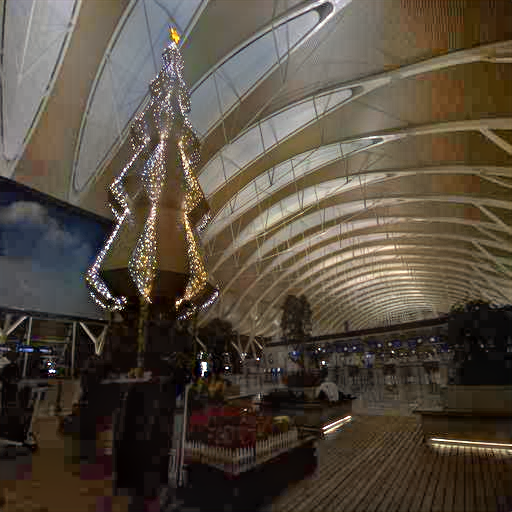} 
    \caption{EnlightenGAN}
    \label{fig:EnlightenGAN_face}
    \end{subfigure}
  \hfill
  \begin{subfigure}{0.19\linewidth}
    \includegraphics[width=1\linewidth]{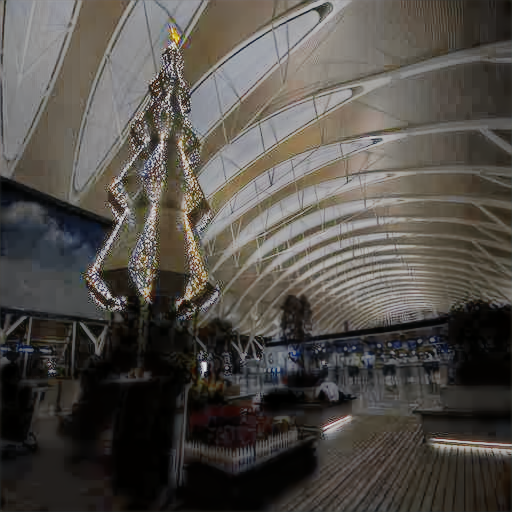} 
    \caption{KinD}
    \label{fig:KinD_face}
    \end{subfigure}
  \hfill
    \begin{subfigure}{0.19\linewidth}
    \includegraphics[width=1\linewidth]{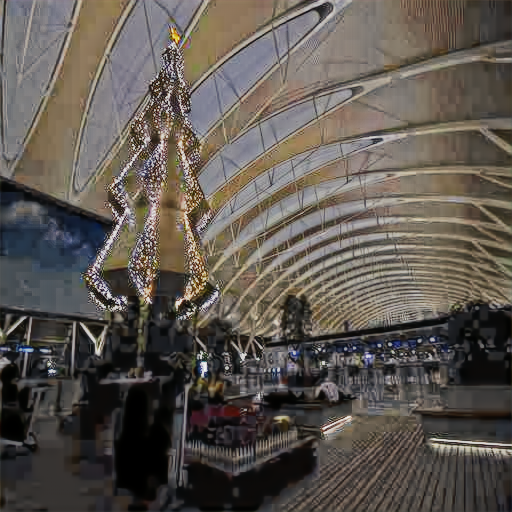} 
    \caption{KinD++}
    \label{fig:KinD++_face}
    \end{subfigure}
  \hfill
    \begin{subfigure}{0.19\linewidth}
    \includegraphics[width=1\linewidth]{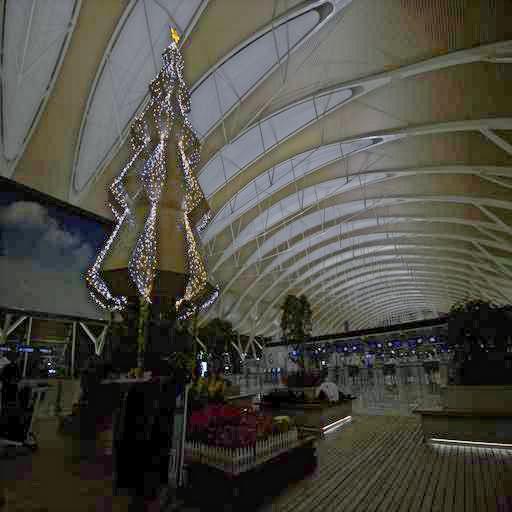} 
    \caption{ZeroDCE}
    \label{fig:dce_face}
    \end{subfigure}
  \hfill
  \begin{subfigure}{0.19\linewidth}
    \includegraphics[width=1\linewidth]{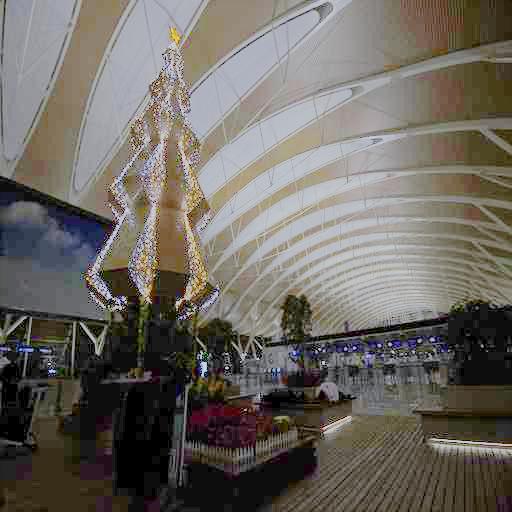} 
    \caption{Ours$^\ast$}
    \label{fig:DRBN_face}
    \end{subfigure}
  \hfill
    \begin{subfigure}{0.19\linewidth}
    \includegraphics[width=1\linewidth]{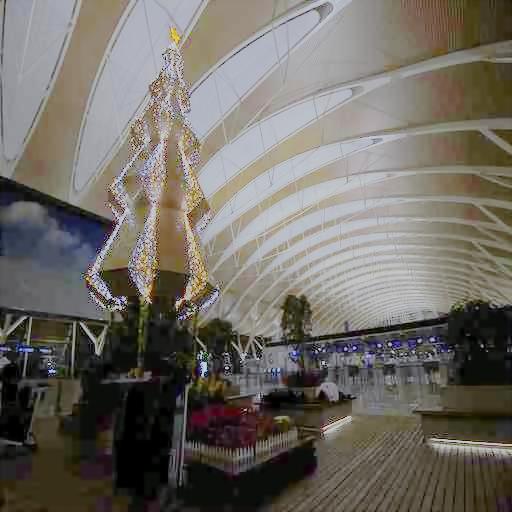} 
    \caption{Ours}
    \label{fig:Ours_face}
    \end{subfigure}
  \caption{Visual comparison for the second one of the 1st series of SCIE$\_$Part2 Dataset.}
  \label{fig:SCIE1}
\end{figure*}

\renewcommand{\thefigure}{A6}
\begin{figure*}[b]
\vspace{-0.3cm}
\setlength{\abovecaptionskip}{0cm} 
\setlength{\belowcaptionskip}{-0.1cm} 
  \centering
  \begin{subfigure}{0.185\linewidth}
    \includegraphics[width=1\linewidth]{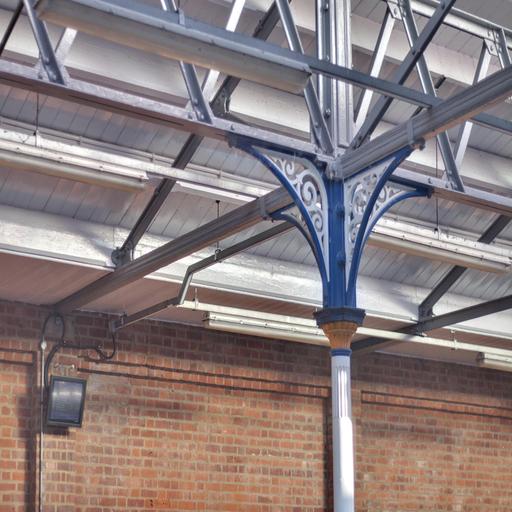} 
    \caption{GT}
    \label{fig:gt}
    \end{subfigure}
  \hfill
    \begin{subfigure}{0.185\linewidth}
    \includegraphics[width=1\linewidth]{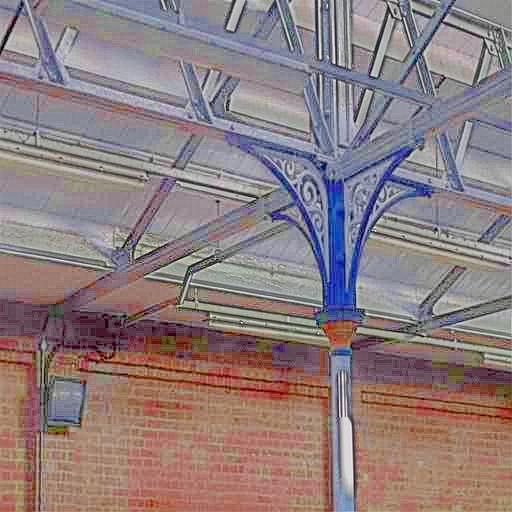} 
    \caption{Retinex-Net}
    \label{fig:Retinex_face}
    \end{subfigure}
  \hfill
    \begin{subfigure}{0.185\linewidth}
    \includegraphics[width=1\linewidth]{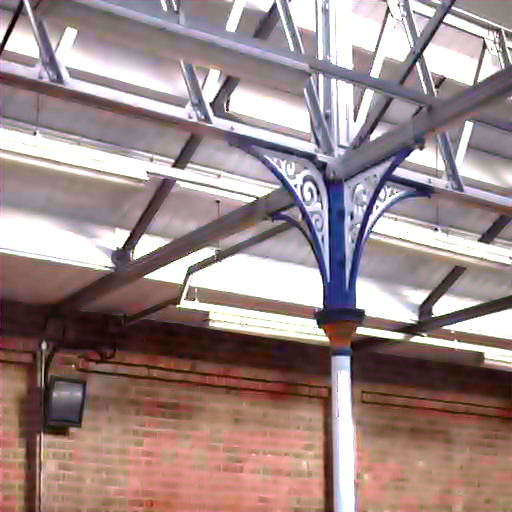} 
    \caption{RUAS}
    \label{fig:RUAS_face}
    \end{subfigure}
  \hfill
  \begin{subfigure}{0.185\linewidth}
    \includegraphics[width=1\linewidth]{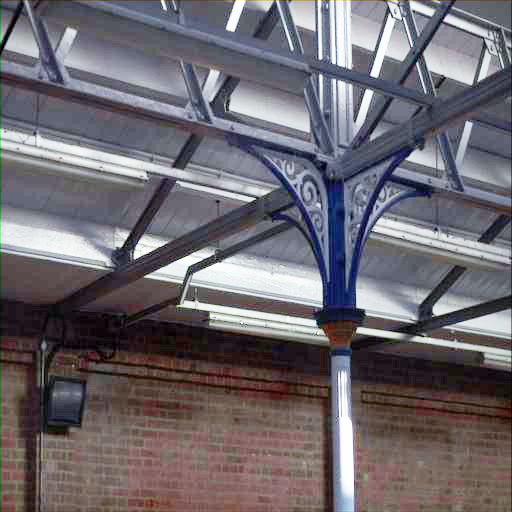} 
    \caption{SCI}
    \label{fig:sci_face}
    \end{subfigure}
  \hfill
  \begin{subfigure}{0.185\linewidth}
    \includegraphics[width=1\linewidth]{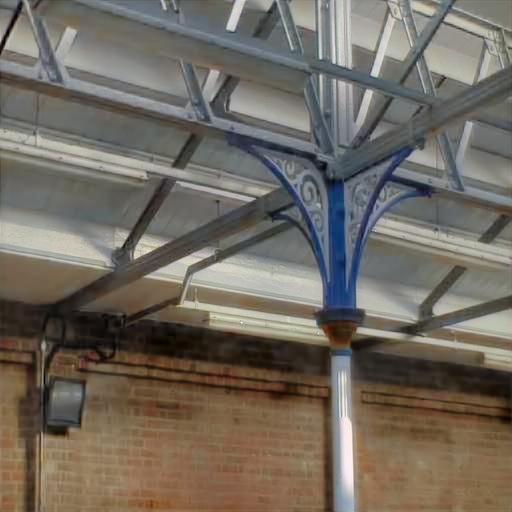} 
    \caption{EnlightenGAN}
    \label{fig:EnlightenGAN_face}
    \end{subfigure}
  \hfill
  \begin{subfigure}{0.185\linewidth}
    \includegraphics[width=1\linewidth]{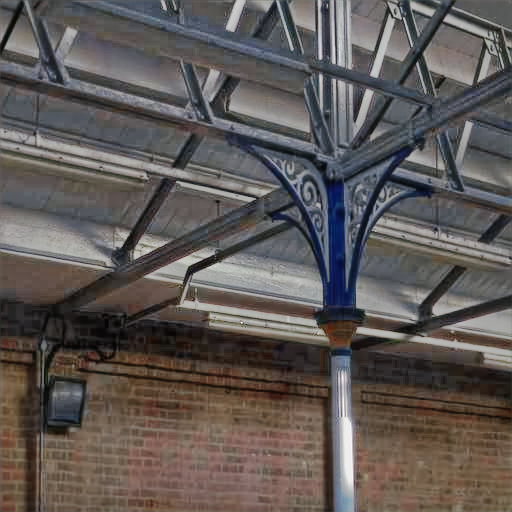} 
    \caption{KinD}
    \label{fig:KinD_face}
    \end{subfigure}
  \hfill
    \begin{subfigure}{0.185\linewidth}
    \includegraphics[width=1\linewidth]{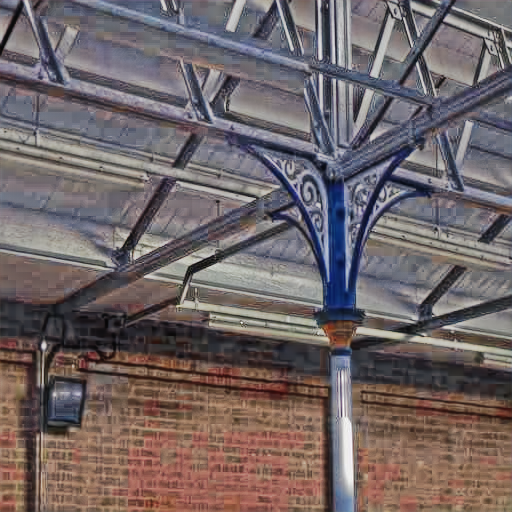} 
    \caption{KinD++}
    \label{fig:KinD++_face}
    \end{subfigure}
  \hfill
    \begin{subfigure}{0.185\linewidth}
    \includegraphics[width=1\linewidth]{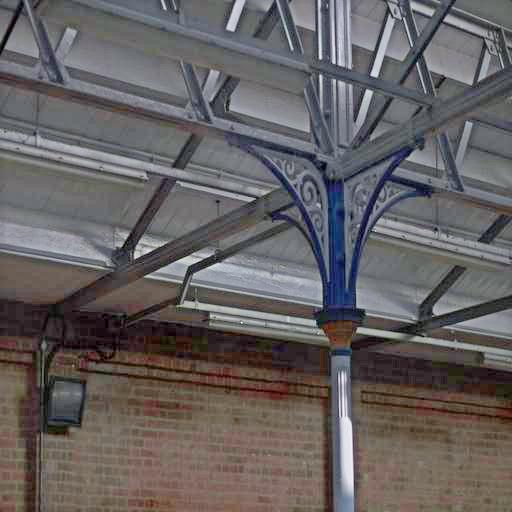}
    \caption{ZeroDCE}
    \label{fig:dce_face}
    \end{subfigure}
  \hfill
  \begin{subfigure}{0.185\linewidth}
    \includegraphics[width=1\linewidth]{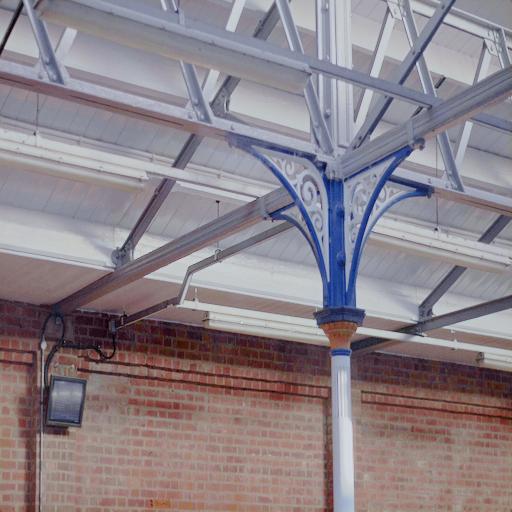} 
    \caption{Ours$^\ast$}
    \label{fig:DRBN_face}
    \end{subfigure}
  \hfill
    \begin{subfigure}{0.185\linewidth}
    \includegraphics[width=1\linewidth]{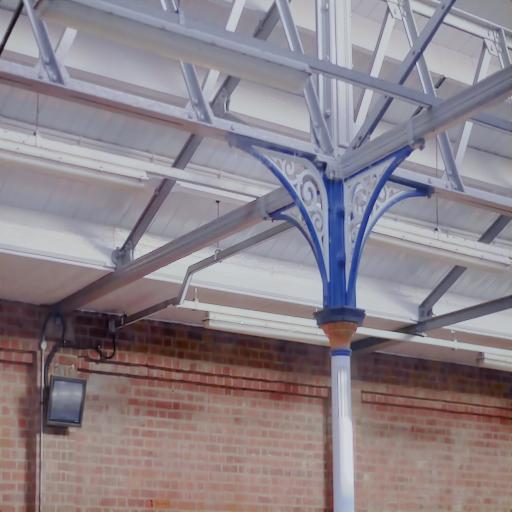} 
    \caption{Ours}
    \label{fig:Ours_face}
    \end{subfigure}
  \caption{Visual comparison for the second one of the 72th series of SCIE$\_$Part2 Dataset.}
  \label{fig:SCIE5}
\end{figure*}
\end{appendices}

\end{document}